\documentclass[10pt,twocolumn,aps,prd,preprintnumbers,superscriptaddress,nofootinbib,amsmath,amssymb,floats,floatfix,showkeys,notitlepage,showpacs,longbibliography]{revtex4-1}
\usepackage{orcidlink}
\usepackage{comment}
\usepackage{lipsum}
\usepackage{graphicx}
\usepackage{subfigure}
\usepackage{palatino}
\usepackage{sans}
\usepackage{hyperref}
\hypersetup{colorlinks=true,linkcolor=blue,urlcolor=blue,citecolor=blue}
\usepackage[toc,page]{appendix}
\usepackage[normalem]{ulem}
\usepackage{adjustbox}
\usepackage{latexsym}
\usepackage{amsmath}
\usepackage{amssymb}
\usepackage{amsfonts}
\usepackage{dcolumn}
\usepackage{bm}
\usepackage{tikz}
\usepackage{bigints}
\usepackage{array,tabularx,multirow,booktabs}
\usepackage[tracking=true]{microtype}
\SetTracking{}{500}
\SetTracking{encoding={*}, shape=sc}{40}
\UseRawInputEncoding 
\allowdisplaybreaks

\begin{document} \sloppy
\title{Analyzing Deflection Angles and Photon Sphere Dynamics of Magnetically Charged Black Holes in Nonlinear Electrodynamic}

\author{Hira Waseem}
\email{hirawaseem633@gmail.com}

\affiliation{Department Of Physics, Quaid-I-Azam University, Islamabad 45320, Pakistan.}
\affiliation{Dipartimento di Fisica, Universit`a degli Studi di Roma "La Sapienza", Piazzale Aldo Moro 5, I-00185 Roma, Italy}

\author{Nikko John Leo S. Lobos \orcidlink{0000-0001-6976-8462}}
\email{nslobos@ust.edu.ph}
\affiliation{Electronics Engineering Department, University of Santo Tomas, Espa\~na Blvd, Sampaloc, Manila, 1008 Metro Manila}

\author{Ali \"Ovg\"un \orcidlink{0000-0002-9889-342X}}
\email{ali.ovgun@emu.edu.tr}
\affiliation{Physics Department, Eastern Mediterranean University, Famagusta, 99628 North
Cyprus via Mersin 10, Turkiye.}

\author{Reggie C. Pantig \orcidlink{0000-0002-3101-8591}}
\email{rcpantig@mapua.edu.ph}
\affiliation{Physics Department, School of Foundational Studies and Education, Map\'ua University, 658 Muralla St., Intramuros, Manila 1002, Philippines}

\date{\today}
\begin{abstract}
In this paper, we investigate the gravitational lensing properties of magnetically charged black holes within the framework of nonlinear electrodynamics. We derive the deflection angle and examine the influence of the nonlinear electrodynamics parameter \(\xi\) on light bending. We initially employ a geometric approach based on the Gauss-Bonnet theorem to analyze the gravitational deflection of null and timelike particles. This method encapsulates the global characteristics of the lensing effect in an elegant manner. In the subsequent part of the work, we explore the impact of nonlinear electromagnetic corrections on the black hole shadow. Using numerical techniques, we study the behavior of the photon sphere and demonstrate that a reduction in the photon sphere radius leads to a correspondingly smaller shadow. We compare these results with those for the Schwarzschild and Reissner-Nordstr\"om black holes, highlighting the distinctive features introduced by nonlinear electrodynamics. Furthermore, we examine the strong deflection limit for light trajectories near these black holes, focusing on the roles of both the magnetic charge \(Q\) and the nonlinear parameter \(\xi\). Our analysis reveals that the combined effects of \(Q\) and \(\xi\) enhance the strong deflection angle, resulting in a more pronounced lensing effect than that predicted by the classical Reissner-Nordstr\"om solution. These findings suggest that the nonlinear interactions may provide a potential observational signature for identifying NED black holes.
\end{abstract}



\maketitle
\section{Introduction} \label{intro}
A long‐standing question in differential geometry has been whether, for a given metric \(g_{\mu\nu}(x)\), one can find a coordinate system in which the metric components are globally constant. This intriguing problem, which captivated many mathematicians in the past \cite{klein}, was resolved by Riemann in 1861 \cite{Reiman}. Riemann showed that any coordinate transformation must preserve the distance between neighboring points, a condition encoded in the Riemann curvature tensor, a fourth-rank tensor essential for characterizing the intrinsic geometry of a manifold. After 1915, Einstein's revolutionary interpretation of gravity transformed our understanding of the phenomenon. Gravity was no longer viewed as a conventional force but as a manifestation of the curvature of spacetime \cite{A.Einstein, Thorne, Rindler}. One of the enduring challenges in modern physics is the unification of quantum field theory with general relativity into a coherent theory of quantum gravity. In relativistic astrophysics, the deflection angle of particles moving in a gravitational field plays a crucial role. When the size of a particle is negligible compared to the gravitational system, its motion can be accurately described as a geodesic, the path representing the shortest distance between two points in curved spacetime. The boundedness or unboundedness of a particle's orbit depends on its kinematical and dynamical parameters, such as energy and angular momentum. Notably, massless particles like photons follow unbounded trajectories, and their paths have been studied extensively.

The bending of light was famously confirmed in 1919 during a solar eclipse by Dyson, Eddington, and Davidson \cite{Dyson}, marking a cornerstone in astrophysics and cosmology as well as a pivotal experimental validation of general relativity. This phenomenon, now known as gravitational lensing, was predicted long before its observation, its early theoretical suggestion even dating back to Solder in 1801. Gravitational lensing not only reveals the presence of massive objects but also provides a powerful means to study exotic structures such as black holes and wormholes \cite{Harada, keeton, Ellis, keeton1, keeton2, chen, sharif, Cao, kogan}. Gravitational lensing is especially versatile because the deflection angle of a light ray is determined by the gravitational field, which itself is governed by the stress-energy tensor of the matter distribution. Once the deflection angle is expressed as a function of the impact parameter, the lensing effect reduces to a geometrical problem. This approach has been employed, for example, to determine the Hubble constant and the cosmic density parameter \cite{roy, refsdal}. In the framework of general relativity, a light ray passing by a spherical mass \(M\) at a minimum distance \(b\) is deflected by the so-called “Einstein angle”
\[
\hat{\alpha} = \frac{4GM}{c^2 b} = \frac{2R_s}{b},
\]
where \(R_s = \frac{2GM}{c^2}\) is the Schwarzschild radius \cite{Bartelmann:1999yn,falco}.

 In 2008, G.~W. Gibbons and M.~C. Werner introduced a novel geometrical approach to gravitational lensing by deriving the deflection angle from the Gaussian curvature of the optical metric. This method exploits the global properties of the deflection angle and employs the Gauss-Bonnet theorem, which elegantly links the intrinsic differential geometry of a surface with its topology. In their pioneering work, Gibbons and Werner applied this technique to analyze the deflection of photons in static, spherically symmetric spacetimes. Subsequently, the same method was adapted to determine the deflection angle for the optical metric of a gravitational lens modeled as a static, spherically symmetric, perfect non-relativistic fluid \cite{Bloomer:2011rd}. In 2012, by utilizing the Randers optical metric, Werner further extended this approach to study the deflection of massless particles, specifically photons in time-independent, axially symmetric (stationary axially symmetric) spacetimes. Werner applied this method to two examples of non-asymptotically flat spacetimes to suggest distance-dependent corrections: one case being the Kottler (Schwarzschild-de Sitter) solution to the Einstein equations, and the other an exact solution in Weyl conformal gravity \cite{werner}. Subsequently, a series of studies have employed the Gauss-Bonnet theorem (GBT) to determine the deflection angle \cite{Godunov:2015nea, Ishihara, Ishihara:2016sfv, a.ovgun, Goulart:2017iko, Ono:2017pie, Fleischer:2017yox, arakida, Ono:2018ybw, Jusufi:2018jof, aliovgun,  ali2, javed, babar, deLeon:2019qnp, Kumar:2019pjp, javed1, aqib, mustafa, Gao:2023sla, Gao:2022cds, Qiao:2022nic, Huang:2022gon, Javed:2023iih, Javed:2022psa,Okyay:2021nnh,Rayimbaev:2022hca}. However, the conventional treatment of boundless orbits, assuming that both the source and observer are located at infinite distances from the lensing mass, has been met with criticism. Some researchers argued that finite distance corrections to the deflection angle should be incorporated when analyzing light propagation in asymptotically curved spacetimes. In response, Rindler and Ishak proposed a definition for the finite distance deflection angle of light in the special case where the lens, receiver, and source are collinear \cite{Rindler:2007zz}. Although this proposal received some criticism \cite{Park:2008ih, Sereno:2008kk, peacock}, a finite--distance deflection angle formulation based on the GW method was later introduced by Ishihara et al. \cite{Ishihara}, which has since gained widespread acceptance; the infinite--distance deflection angle can then be recovered as a limiting case. More recently, Li et al. employed the Jacobi-Maupertuis-Randers-Finsler metric in conjunction with the GB theorem to explore finite distance effects on the deflection of heavy particles in time-independent, axially symmetric spacetimes \cite{Li:2020zxi, ATLAS:2024mih, Li:2020dln}. Additionally, Arakida examined the deflection angle in the finite region by considering a vacuum solution \cite{Arakida:2017hrm}, and Takizawa et al. investigated light deflection in asymptotically curved spacetimes using non-inertial circular orbits \cite{Takizawa:2020egm}. In this research, a geometric approach is employed to first derive the radius of the circular orbit for a particle, and then the Gauss-Bonnet theorem is applied to calculate its deflection angle.

 Pioneering studies on this phenomenon were conducted by Jean-Pierre Luminet in the late 1970s \cite{Luminet:1979nyg}, who established the theoretical framework for visualizing black hole shadows by demonstrating how gravitational bending of light produces such distinct silhouettes. This theoretical foundation was spectacularly confirmed by the Event Horizon Telescope collaboration, which in 2019 captured the first image of a black hole shadow in the M87 galaxy \cite{EventHorizonTelescope:2019dse,EventHorizonTelescope:2019ths,EventHorizonTelescope:2019ggy} and later in the Milky Way galaxy in 2022 \cite{EventHorizonTelescope:2022wkp,EventHorizonTelescope:2022wok,EventHorizonTelescope:2022xqj}. Such empirical evidence not only validates general relativity under extreme gravitational conditions but also marks a pivotal moment in observational astrophysics.

The study of black hole shadows is crucial for several reasons. It provides a direct test of gravitational theories in regions of intense curvature, offers insights into the behavior of matter (and even dark matter or other cosmological background parameters) as well as radiation near the event horizon \cite{Yan:2024rsx,Yang:2024ulu,Yang:2023tip,Chakhchi:2024tzo,Jafarzade:2024knc,Ovgun:2023wmc,Pantig:2024rmr,Nozari:2024vxp,Sui:2023tje,Zare:2024dtf,Pantig:2022sjb,Sucu:2024xck}, and enables the precise measurement of key black hole parameters such as mass, spin, and those emerging from alternative theories of gravity \cite{AraujoFilho:2024xhm,AraujoFilho:2024lsi,Uniyal:2023inx,KumarWalia:2024yxn,Uniyal:2023ahv,Lambiase:2024uzy,Pantig:2024ixc,Pantig:2024kqy,Li:2024owp,Khodadi:2024ubi,Ali:2024cti,Meng:2024puu,Heidari:2023egu,Heidari:2023bww,Calza:2024xdh,Calza:2024fzo}. Consequently, black hole shadows serve as a unique diagnostic tool in probing the fundamental physics of gravity in its most extreme regimes.

Gravitational lensing in the strong deflection limit, where light trajectories pass extremely close to the photon sphere, provides a powerful means to probe the interplay between electromagnetic fields and gravity in high-energy astrophysical environments \cite{Bozza:2002zj, Tsukamoto:2016jzh, Tsukamoto:2016oca, Fu:2021akc, Soares:2023err, Soares:2023uup}. Although the weak deflection angle has been extensively studied in various black hole spacetimes, the strong deflection regime is particularly relevant for near-horizon physics, where deviations from general relativity become most pronounced. In the case of nonlinear electrodynamic (NED) magnetic black holes, the presence of nonlinear electromagnetic fields alters both the photon sphere radius and the critical impact parameter, leading to measurable differences in light bending.

 \textcolor{black}{While numerous nonlinear electrodynamics (NED) theories, including Born-Infeld (BI), Euler-Heisenberg (EH), and logarithmic electrodynamics, have been demonstrated to influence photon spheres and black hole shadows, the majority of these studies rely on perturbative or numerical approaches. For instance, the Born-Infeld model admits an analytically expressible magnetically charged solution whose shadow radius increases with the Born-Infeld nonlinearity parameter \cite{He:2022opa}. Euler-Heisenberg and Bronnikov-type NED solutions have been utilized to place observational constraints on loop-correction terms and magnetic charges by comparing theoretical predictions with the Event Horizon Telescope (EHT) measurements of the M87$^*$ black hole shadow \cite{Allahyari:2019jqz}. The double-logarithmic NED model proposed by Gullu and Mazharimousavi yields an exact magnetically charged solution expressible in elementary functions, enabling detailed thermodynamic analysis; however, explicit computations of the shadow were not performed \cite{Gullu:2020qni}. Kruglov introduced an exponential NED theory leading to regular black hole solutions characterized by horizon structures involving incomplete gamma functions, exhibiting distinctive thermodynamic phase transitions \cite{Kruglov:2018rrm}. Additionally, Ma explored magnetically charged regular black hole solutions in another class of NED models \cite{Ma:2015gpa}, and Dymnikova’s electrically charged regular black hole solutions were recently found to produce shadows approximately $10\%$ larger than their Reissner-Nordstr\"om counterparts \cite{Dymnikova:2021dqq}. Despite these advancements, an entirely analytic magnetically charged NED black hole solution, maintaining a purely Schwarzschild-like asymptotic form and providing closed-form expressions for weak and strong deflection angles, photon sphere radii, and shadow observables, remains absent from the literature.}

 \textcolor{black}{In this work, we bridge this gap by employing the exact magnetically charged solution recently introduced by Mazharimousavi \textit{et al.} \cite{Mazharimousavi:2023omd}. Using this analytical framework, we derive closed-form expressions for the weak deflection angle using the Gauss–Bonnet theorem and perform a comprehensive strong deflection limit analysis. We further compute explicit photon sphere radii and shadow radii as functions of the NED coupling parameter $\xi$, facilitating direct quantitative comparisons with Schwarzschild and Reissner-Nordstr\"om spacetimes. Finally, by confronting our theoretical results with EHT observational data of M87$^*$ and Sgr $A^*$, we establish new observational constraints on the NED coupling parameter. These combined analytical and observational features distinguish our contribution from previous works, providing a comprehensive and self-consistent treatment of magnetically charged NED black holes.}

This paper is organized as follows. In Section II, we review the magnetically charged black holes in nonlinear electrodynamics, then in Section III, we introduce the geometric approach for understanding spacetime geodesics and detail the application of the Gauss-Bonnet theorem to derive the gravitational deflection angle. In Section IV, we calculate the strong deflection angle of photons. Section~V is devoted to calculating the photon sphere radius and shadow radius. Finally, in Section VI, we summarize our findings.

\section{Brief Review of Magnetically Charged Black Holes in Nonlinear Electrodynamics} \label{sec2}
The action of Einstein's non-linear electrodynamics (NED), which is the linearized Maxwell's theory in the weak field domain, written as,
\begin{equation}
    S = \int d^{4}x \sqrt{-g} \left( \frac{\mathcal{R}}{16\pi} + \mathcal{L}\left(\mathcal{F}\right) \right), \label{eq:2}
\end{equation}
where $\mathcal{R}$ stands for the Ricci scalar, $G=1$ (Gravitational constant) and $\mathcal{L}\left(\mathcal{F}\right)$ is given as \cite{Mazharimousavi:2023omd}
\begin{equation}
    \mathcal{L}=-\varepsilon\left(\mathcal{F}\right)\mathcal{F}, \label{eq:3}
\end{equation}
and 
\begin{equation}
    \varepsilon\left(\mathcal{F}\right) = \frac{16\left(3\sqrt{2\mathcal{F}} + \xi\left(\xi + \sqrt{\xi^2 + 4\sqrt{2\mathcal{F}}}\right)\right)}{3\left(\xi + \sqrt{\xi^2 + 4\sqrt{2\mathcal{F}}}\right)^4}. \label{eq:4}
\end{equation}

Note that $\xi$ represents the positive constant parameter. If $\xi$ is small, then,
\begin{equation}
    \varepsilon\left(\mathcal{F}\right)  \simeq 1-\frac{4}{3}\frac{\sqrt[4]{2}}{\sqrt[4]{\mathcal{F}}}\xi+\left(\frac{\sqrt[4]{2}}{\sqrt[4]{\mathcal{F}}}\right)^{2}\xi^2+\mathcal{O}\left(\xi^{3}\right). \label{eq:5}
\end{equation}
where this $\mathcal{F}$ is electromagnetic tensor, $\frac{1}{4}F_{\mu \nu}F^{\mu \nu}$.
Einstein NED equation takes the following form after the variation of action, Eq. (\ref{eq:2}) w.r.t NED energy-momentum tensor,
\begin{equation}
    G^{\nu}_{\mu}=8\pi T^{\nu}_{\mu},\label{eq:6}
\end{equation}
where the energy-momentum tensor $T^{\nu}_{\mu}$ is 
\begin{equation}
    T_{\mu}^{\nu}=\frac{1}{4\pi}\left(\mathcal{L}\delta_{\mu}^{\nu}-\mathcal{L}_{\mathcal{F}}\mathcal{F}_{\mu \lambda}\mathcal{F}^{\nu \lambda}\right). \label{eq:7}
\end{equation}
 Maxwell's NED equations are yielded in the result of variation of action w.r.t gauge fields written as
\begin{equation}
    d\left(\mathcal{L}_{\mathcal{F}} \tilde{\mathbf{F}}\right) = 0. \label{eq:8}
\end{equation}

where this $\tilde{\mathbf{F}}$ is the dual electromagnetic field tensor,
\begin{equation}
    \textbf{F}=\frac{1}{2}F_{\mu \nu }dx^{\mu} \wedge dx^{\nu}. \label{eq:9}
\end{equation}
Considering a magnetic monopole located at the origin with the field
 to form,
\begin{equation}
    \textbf{F}=Q \,\sin\theta \, d\theta \wedge d\phi. \label{eq:10}
\end{equation}
The tt component of Eq. (\ref{eq:7}) gives,
\begin{equation}
    \frac{rf^{\prime}\left(r\right)+f\left(r\right)-1}{r^{2}}=-\frac{4B^{3}\left(2\xi^{2}+2\xi\sqrt{\xi^{2}+2B}+3B\right)}{3\left(\xi+\sqrt{\xi^{2}+2B}\right)^{4}}. \label{eq:11}
\end{equation}
where $B=\frac{Q}{r^{2}}$ is the radial component of the magnetic field of the magnetic pole $Q>0$.
The exact solution of the Eq. (\ref{eq:11}) is \cite{Mazharimousavi:2023omd}, 
\begin{equation}
\begin{split}
f(r) &= 1 - \frac{2M}{r} + \frac{Q^2}{r^2}
         - 2Q\,\xi^2
         - \frac{2\,\xi^4\,r^2}{9} \\[6pt]
     &\quad - \frac{4\sqrt{2}\,Q^{3/2}\,\xi}{3r}
         \ln\!\Biggl(
           \frac{4Q + 2\sqrt{2Q}\,\sqrt{\xi^2r^2 + 2Q}}{r}
         \Biggr) \\[6pt]
     &\quad + \frac{2\,\xi\,\sqrt{\xi^2r^2 + 2Q}\,\bigl(\xi^2r^2 + 8Q\bigr)}{9r}
\end{split}
\label{eq:12}
\end{equation}
The metric function satisfies all the Einstein field equations. This metric function for the limit $\xi \to 0$ gives the standard Reissner Nordstr\"om black hole
\begin{equation}
    f\left(r\right)=1-\frac{2M}{r}+\frac{Q^{2}}{r^{2}}, \label{eq:13}
\end{equation}
while the exact behavior of $f\left(r\right)$ in the small $\xi$ domain is,
\begin{equation}
    f\left(r \right)=1-\frac{2M}{r}+\frac{Q^{2}}{r^{2}}+\xi \left(\frac{4Q^{3/2}\sqrt{2}\left(4-3\text{ln}\left(\frac{8Q}{r}\right)\right)}{9r}\right)+\mathcal{O}\left(\xi^{2}\right). \label{eq:14}
\end{equation}

\begin{figure}
    \centering
    \includegraphics[width=0.5\textwidth]{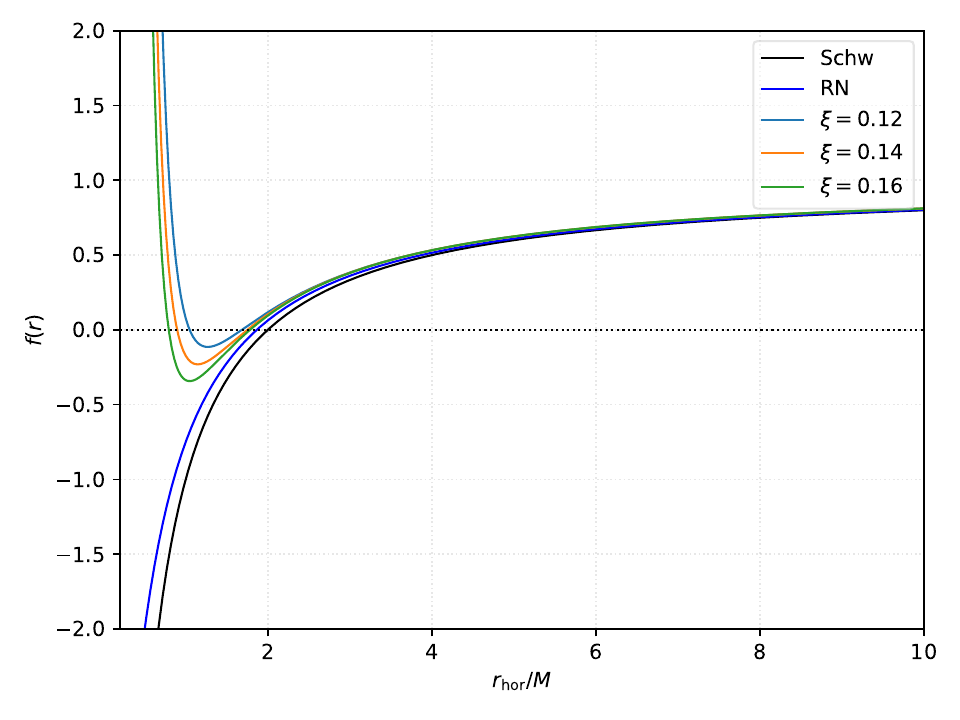}
    \caption{Behavior of the metric function $f(r)$. Choosing a very small value for $\xi$ makes the horizon to disappear. Here, we have chosen an arbitrary value of $Q/M = 0.50$ to enable comparison.}
    \label{fig:1a}
\end{figure}

Fig. (\ref{fig:1a}) depicts the behavior of the metric function, as compared to the Schwarzschild and Reissner-Nordstr\"om BHs.  The plot of the lapse function \( f(r) \) reveals important properties of the horizons for the magnetically charged black hole. The event horizons are identified where \( f(r) = 0 \), and the presence of multiple horizons suggests deviations from the simple Schwarzschild case. As the parameter \( \xi \) increases, the outer horizon shifts inwards. Compared to the RN and Schwarzschild solutions, the modified black hole exhibits a richer horizon structure, potentially affecting its stability and causal properties. 

We now turn our attention to the thermodynamic properties of the black hole solution given by the metric function given in Eq. \ref{eq:14}. The horizon radius \( r_h \) is defined by the condition \( f(r_h)=0 \). Solving this condition explicitly for the ADM mass \( M \), we obtain:
\begin{equation}
    M(r_h)=\frac{r_h}{2}+\frac{Q^2}{2r_h}+\xi\frac{2\sqrt{2}\,Q^{3/2}\left[4-3\ln\left(\frac{8Q}{r_h}\right)\right]}{9}.
    \label{eq:mass_horizon}
\end{equation}
Equation~(\ref{eq:mass_horizon}) clearly shows corrections induced by the coupling parameter \( \xi \). For \( \xi=0 \), the expression reduces exactly to the well-known Reissner-Nordstr\"om mass-horizon relationship.

The Hawking temperature \( T_H \), obtained from the surface gravity at the horizon, is defined as:
\begin{equation}
    T_H=\frac{1}{4\pi}\frac{df(r)}{dr}\Bigg|_{r=r_h}=\frac{1}{4\pi}\left[\frac{1}{r_h}-\frac{Q^2}{r_h^3}+\xi\frac{4\sqrt{2}\,Q^{3/2}}{3r_h^2}\right].
    \label{eq:temp_hawking}
\end{equation}
This temperature shows explicitly how the coupling \( \xi \) modifies the black hole thermodynamics, adding a nontrivial logarithmic dependence implicitly through the horizon radius \( r_h \).

We next compute the heat capacity \( C \), an essential quantity to assess thermodynamic stability, defined by:
\begin{equation}
    C=\frac{dM/dr_h}{dT_H/dr_h}=4\pi\frac{\frac{r_h^4}{2}-\frac{Q^2r_h^2}{2}+\xi\frac{2\sqrt{2}\,Q^{3/2}r_h^3}{3}}{-r_h^2+3Q^2-\xi\frac{8\sqrt{2}\,Q^{3/2}r_h}{3}}.
    \label{eq:heat_capacity}
\end{equation}
\begin{figure}
    \centering
\includegraphics[width=1.1\linewidth]{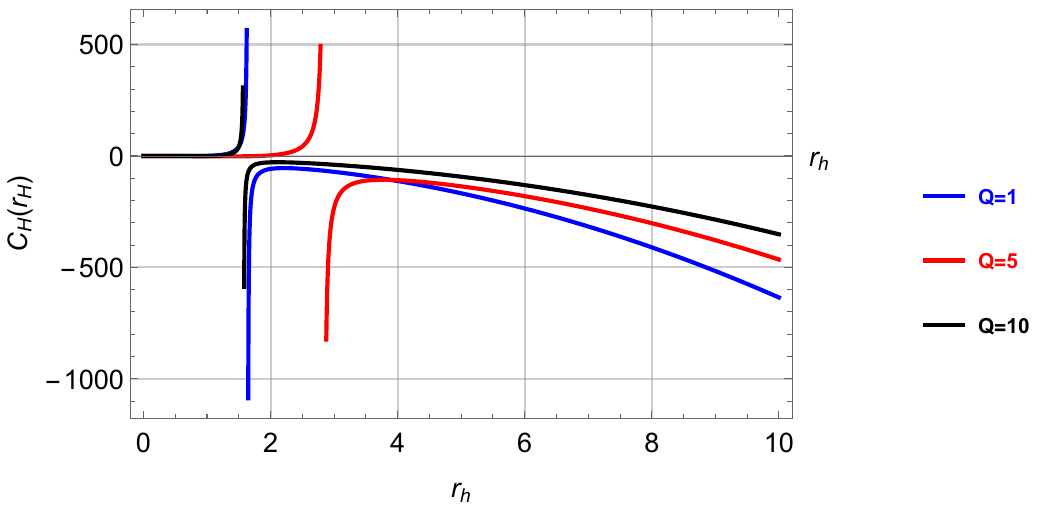}
    \caption{\textcolor{black}{Heat capacity} for different values of charge for $\xi=0.10$}
    \label{fig:heat}
\end{figure}

The heat capacity (Eq.~(\ref{eq:heat_capacity})) is critical for evaluating thermodynamic stability. Specifically, the sign of \( C \) dictates stability:
\begin{itemize}
    \item \( C>0 \): Black hole configuration is thermodynamically stable. Small perturbations in the mass result in restoring temperature fluctuations, thus stabilizing the system.
    \item \( C<0 \): Configuration is unstable. Fluctuations grow without bound, leading to evaporation or runaway instabilities.
\end{itemize}

Analyzing Eq.~(\ref{eq:heat_capacity}), it is clear that the stability behavior depends intricately on the interplay between the horizon radius \( r_h \), charge \( Q \), and the coupling parameter \( \xi \). For instance, in the standard Reissner-Nordstr\"om limit (\(\xi=0\)), stability conditions reduce to the known criterion, and thermodynamic instability occurs typically for smaller black holes. However, the coupling \(\xi\) introduces new stability regions, potentially stabilizing previously unstable configurations or vice versa, depending explicitly on its magnitude and sign.

On the other hand, the asymptotic behavior of this metric function is intriguing when the following form is considered:
\begin{equation}
    f\left(r\right)=1-\frac{2M_{\rm ADM}}{r}+\frac{Q^{3}}{9\xi^{2}r^{4}}+\mathcal{O}\left(\frac{1}{r^{6}}\right), \label{eq:15}
\end{equation}

where,
\begin{equation}
    M_{\rm ADM} = M + \frac{2\sqrt{2}}{3} \xi Q^{3/2} \ln\left(2\xi \sqrt{2Q}\right). \label{eq:16}
\end{equation}

\section{Calculation of the weak field deflection angle} \label{sec3}

{\color{black}
In this section, we investigate the deflection angle of particles in magnetically charged black holes within nonlinear electrodynamics. We employ the Gauss-Bonnet theorem applied to the optical metric to derive the weak-field deflection angle.

The Gauss-Bonnet theorem relates a surface's intrinsic curvature to its topology \cite{carmo}.  Using the Gibbons and Werner method given in \cite{Gibbons:2008rj}, we will calculate the weak deflection angle.

Consider a static, spherically symmetric spacetime described by the line element:
\begin{equation}
ds^2 = -A(r)dt^2 + B(r)dr^2 + C(r)d\Omega^2. \label{eq:17}
\end{equation}

The corresponding optical metric is written as: 
\begin{equation}
dt^2 = \frac{B(r)}{A(r)}\,dr^2 + \frac{C(r)}{A(r)}\,d\phi^2.
\end{equation}

The photon trajectory follows \cite{Li:2019vhp}:
\begin{equation}
\left(\frac{du}{d\phi}\right)^{2} = \frac{C(r)^4 u^4}{A(r)B(r)}\Biggl[\frac{1}{b^2} + \frac{A(r)}{C(r)}\Biggr],
\quad u\equiv\frac{1}{r},
\end{equation}
where $b$ is the impact parameter. The photon orbit equation becomes
\begin{equation}
    \left(\frac{du}{d\phi}\right)^2\;=\;\frac{1}{b^2}-u^2+2M_{\rm ADM}u^{3}-\frac{Q^{3}u^{6}}{3\xi^{2}}. \label{eq:50}
\end{equation}

whose perturbative solution reads:
\begin{equation}
\begin{split}
u &= \frac{\sin\phi}{b}
     + \frac{M_{\rm ADM}\bigl(1 + \cos^2\!\phi\bigr)}{b^2} \\[4pt]
  &\quad - \frac{80\,Q^{3}\,\sin\phi}{384\,b^{5}\,\xi^{2}}
     + \mathcal{O}\!\bigl(M_{\rm ADM}^{2},\,\xi^{2},\,(Q^{3})^{2}\bigr).
\end{split}
\label{eq:51}
\end{equation}

Then we calculate the Gaussian curvature using $\mathcal{K}=R_{icciScalar}/2$ as follows:
\begin{equation}
\begin{split}
\mathcal{K}
&= \frac{M_{\rm ADM}\bigl(3M_{\rm ADM}-2r\bigr)}{r^{4}}
  + \frac{2Q^{6}}{27\,r^{10}\,\xi^{4}}
  - \frac{2Q^{3}\bigl(9M_{\rm ADM}-5r\bigr)}{9\,r^{7}\,\xi^{2}} \\[6pt]
&= \frac{3M^2}{r^4}
  - \frac{2M}{r^3}
  + \frac{2Q^6}{27\,r^{10}\,\xi^4}
  - \frac{2M\,Q^3}{r^7\,\xi^2}
  + \frac{10\,Q^3}{9\,r^6\,\xi^2} \\[6pt]
&\quad
  - \frac{4\sqrt{2}\,Q^{9/2}\,\ln\!\bigl(2\sqrt{2}\sqrt{Q}\,\xi\bigr)}{3\,r^7\,\xi}
  + \frac{4\sqrt{2}\,M\,Q^{3/2}\,\xi\,\ln\!\bigl(2\sqrt{2}\sqrt{Q}\,\xi\bigr)}{r^4} \\[4pt]
&\quad
  - \frac{4\sqrt{2}\,Q^{3/2}\,\xi\,\ln\!\bigl(2\sqrt{2}\sqrt{Q}\,\xi\bigr)}{3\,r^3}
  + \frac{8\,Q^3\,\xi^2\,\ln^2\!\bigl(2\sqrt{2}\sqrt{Q}\,\xi\bigr)}{3\,r^4}\,.
\end{split}
\label{eq:55}
\end{equation}

For asymptotically flat spacetimes, the deflection angle can be written as:
\begin{eqnarray}
\alpha&=& \int\mathcal{K}\sqrt{det\left(g_{opt}\right)}\,dr\,d\phi.  \label{eq:56}
\end{eqnarray}
Expanding in the weak-field limit gives, the deflection angle is calculated as follows:
\begin{equation}
\begin{split}
\alpha 
&= \frac{4M}{b}
  - \frac{5\pi\,Q^3}{48\,b^4\,\xi^2}
  + \frac{\sqrt{2}\,M\,\pi\,Q^{3/2}\,\xi\,
      \ln\!\bigl(2\sqrt{2}\sqrt{Q}\,\xi\bigr)}{b^2} \\[6pt]
&\quad
  + \frac{64\,M\,Q^3\,\xi^2\,
      \ln^2\!\bigl(2\sqrt{2}\sqrt{Q}\,\xi\bigr)}{9\,b^3}
  + \frac{8\sqrt{2}\,Q^{3/2}\,\xi\,
      \ln\!\bigl(2\sqrt{2}\sqrt{Q}\,\xi\bigr)}{3\,b} \\[6pt]
&\quad
  + \frac{2\pi\,Q^3\,\xi^2\,
      \ln^2\!\bigl(2\sqrt{2}\sqrt{Q}\,\xi\bigr)}{3\,b^2}
  + \mathcal{O}\!\bigl(M^2,\,\xi^2,\,(Q^3)^2\bigr).
\end{split}
\label{eq:57}
\end{equation}

This deflection angle expression encapsulates both the standard Schwarzschild term, \( \frac{4M}{b} \), and additional corrections arising from the modified metric $A(r)$
where the extra terms, featuring both power-law and logarithmic dependences on the impact parameter \(b\), reflect non-trivial modifications due to the charge-like parameter \(Q\) and the coupling \(\xi\). These corrections, which become more pronounced for smaller \(b\), hint at richer underlying physics—-such as non-linear electrodynamics or higher-curvature effects-and suggest observable deviations in gravitational lensing, with the expansion being perturbative in \(M\), \(\xi\), and \(Q^3\).
For Schwarzschild spacetime, the deflection angle can be obtained from the above expression by taking $Q \to 0$ and $\xi \to 0$, that is, $\frac{4M}{b}$.

A graphical depiction of the deflection angle for different values of $\xi$ is shown in Fig.~\ref{fig:3}. This plot illustrates the gravitational deflection angle \( \alpha \) as a function of the impact parameter \( b \) for a magnetically charged black holes in nonlinear electrodynamics with mass \( M = 1 \) and charge \( Q = 0.3 \), analyzed under different values of the parameter \( \xi \). The results indicate that for small impact parameters, the deflection angle exhibits a sharp peak, with the magnitude of bending increasing as \( \xi \) grows. This suggests that \( \xi \) enhances the gravitational lensing effect, potentially modifying the spacetime geometry and amplifying light bending. As \( b \) increases, the deflection angle decreases and converges toward zero, consistent with weak-field gravitational lensing expectations. These deviations from standard Reissner-Nordstr\"om behavior could have observational implications for black hole lensing and shadow analysis, providing potential tests for modified gravity models.

\begin{figure}
    \centering
\includegraphics[width=0.5\textwidth]{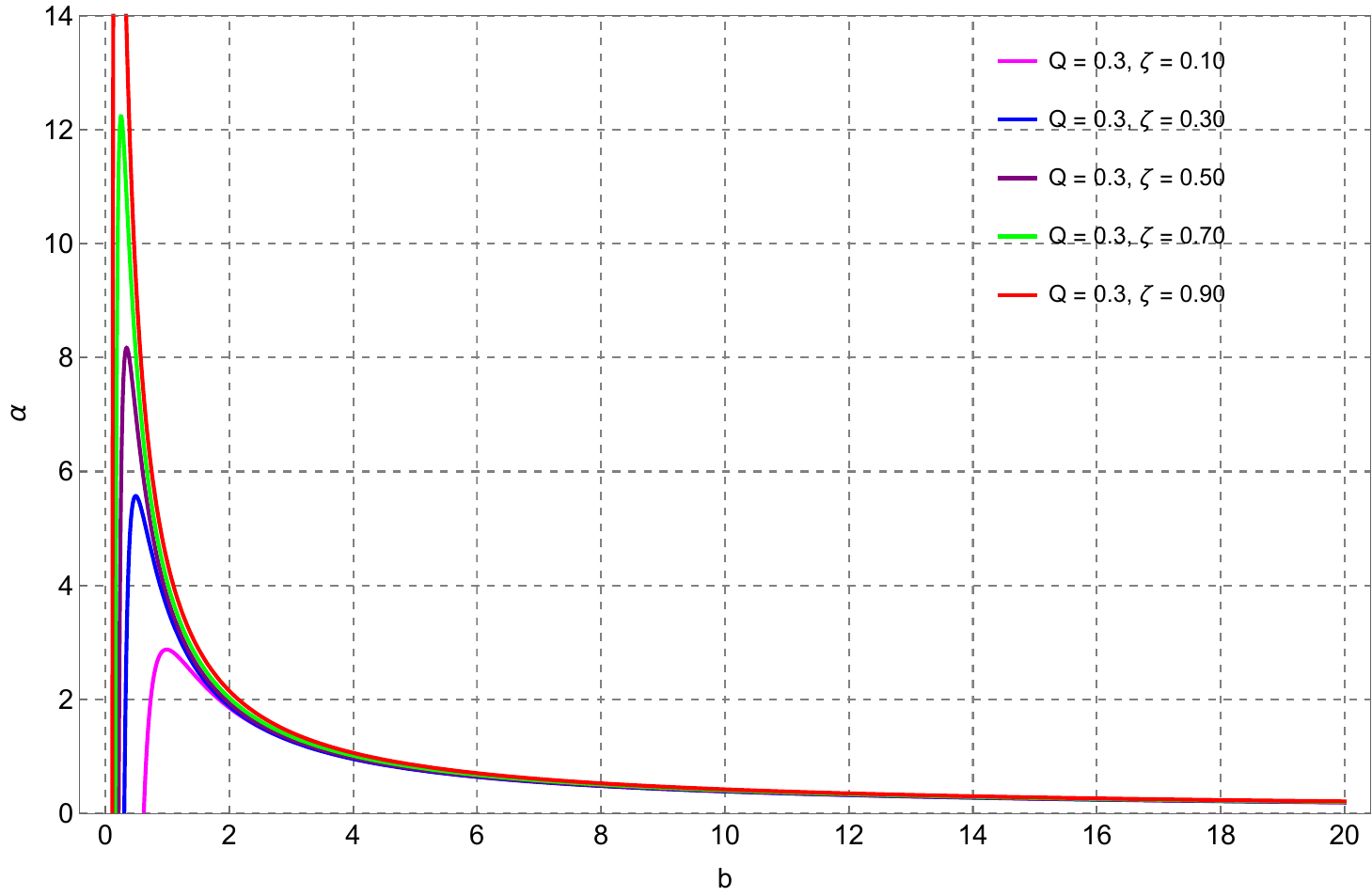}
    \caption{Behavior of the weak deflection angle $\alpha$. Here, we kept $M=1$, $Q=0.3M$.}
    \label{fig:3}
\end{figure}
When light just passes closest to the source, then the post-Newtonian (PPN) formalism equation for light deflection is written as \cite{Will:2018bme},
\begin{eqnarray}
    \alpha=175^{\prime\prime}\left(\frac{1+\gamma}{2}\right).\label{g5}
\end{eqnarray}
where $\gamma$ is the PPN deflection parameter \cite{Filchenkov:2018wom}. By equating Eq. (\ref{eq:57}) and Eq. (\ref{g5}), the constraint on the RN scale can be determined easily, which comes out to be:
\begin{eqnarray}
    0<\xi<4.7\times 10^{-2}.\label{g6}
\end{eqnarray}

 The tight bound on $\xi$ has significant implications. Many alternative theories of gravity predict modifications to the standard gravitational potential, and these modifications often manifest in observable quantities such as the deflection of light. This result implies that any extra contributions to the gravitational potential, parameterized by $\xi$, must be very small to remain consistent with the observed light deflection. In effect, the observational success of GR in explaining light bending forces the RN-scale modifications (or analogous corrections) to lie within a narrow range. Any theory predicting a larger value of $\xi$ would lead to a deflection angle in conflict with high-precision solar system measurements. Thus, this relation acts as an important filter for theoretical models, ruling out those that cannot accommodate the smallness of $\xi$.

\section{Calculation of the strong deflection of angle photons} \label{sec4}
Investigating the effects of black holes on the photons in the strong-field region, \textcolor{black}{we adopt the methodology outlined by Tsukamoto} in Ref. \cite{Tsukamoto:2016jzh}. \textcolor{black}{The strong deflection angle quantifies the bending of light when passing very close to a black hole. It is derived from the orbit equation expressed as},

\begin{equation}
\label{eq.43}
\left(\frac{dr}{d\phi}\right)^{2} = \frac{\mathcal{R}(r)r^2}{B(r)},
\end{equation}
where the radial function is provided as,
\begin{equation}
\label{eq.44}
\mathcal{R}(r) = \frac{A({r_0})r^2}{A(r)r_{0}^2}-1.
\end{equation}

\textcolor{black}{It is noted that $A(r)$ the metric function defined by Eq. \eqref{eq:15} and \eqref{eq:16}, evaluating the metric function $A(r)$ at distance, $r_{0}$, yields $A(r_{0})$. It is essential for relating the deflection angle to the closest approach the light can make near the black hole. The solution of Eq. \eqref{eq.43} yields the strong deflection angle $\alpha(r_{0})$ as shown in Ref. \cite{Tsukamoto:2016jzh, Bozza:2002zj},}
\begin{equation}
\begin{split}
\label{eq.45}
\alpha(r_{0}) &= I(r_{0}) - \pi \\
&= 2 \int^{\infty}_{r_{0}} \frac{dr}{\sqrt{\frac{\mathcal{R}(r)C(r)}{B(r)}}}-\pi.
\end{split}
\end{equation}
\textcolor{black}{The integral in Eq. \eqref{eq.45} cannot be analytically evaluated; hence we resort to analytic approximation employing a series expansion over $r=r_{0}$ resulting to an integration that is composed of regular integral and diverging integral as shown in Ref. \cite{Tsukamoto:2016jzh}. The details of the expansion of Eq. \eqref{eq.45} were shown in Refs. \cite{Tsukamoto:2016jzh, Bozza:2002zj}. Substituting the new variable, $z \equiv 1-\frac{r_{0}}{r}$ and evaluating the integral finally provides the strong deflection angle,}
\begin{equation}
\label{eq.48}
\hat{\alpha}_{\text{str}} = -\bar{a} \log \left(\frac{b_0}{b_\text{crit}}-1\right)+\bar{b}+O\left(\frac{b_{0}}{b_{c}}-1\right)\log\left(\frac{b_{0}}{b_{c}}-1\right),
\end{equation}
where $\bar{a}$ and $\bar{b}$ are constant values contributing to deflection angle. While, $b_{0}$ and $b_\text{crit}$  are the impact parameter evaluated at the closest approach, $r_{0}$, and critical impact parameter, respectively. \textcolor{black}{The first term in Eq. \eqref{eq.48} arises from the diverging integral. It indicates that as the impact parameter approaches the critical impact parameter, the entire expression diverges to infinity. In this regime, light is captured by the photon sphere, becoming trapped in its orbit. The second term, which results from the regular integral, contributes to the overall deflection angle. In certain strong deflection angle (SDA) expressions, such as that of the Reissner-Nordstr\"om (RN) black hole, the charge influences the regular integral, introducing a logarithmic correction to the deflection angle.  The coefficients $\bar{a}$ and $\bar{b}$ are expressed as \cite{Tsukamoto:2016jzh},}
\begin{equation}
\label{abar}
\bar{a} = \sqrt{\frac{2B(r_\text{ps})A(r_\text{ps})}{2A(r_\text{ps}) - A''(r_\text{ps})r_\text{ps}^2}},
\end{equation}
and
\begin{equation}
\label{logarg}
\bar{b} = \bar{a} \log\left[r_\text{ps} \left( \frac{2}{r_\text{ps}^2}-\frac{A''(r_\text{ps})}{A(r_\text{ps})}\right) \right]+I_{R}(r_\text{ps})-\pi,
\end{equation}
where $A(r_\text{ps})$ is metric function evaluated at the photon sphere, and $I_{R}$ is the regular integral evaluated from 0 to 1. \textcolor{black}{This change of integration limit is due to the fact we integrate in terms of the variable $z$ to retain the $r_\text{ps}$ limit in the expression.} The double prime in Eq. \eqref{abar} and Eq. \eqref{logarg} correspond to the second derivative with respect to $r$ evaluated on $r_\text{ps}$.

The metric functions presented above are derived through a series expansion. Unlike the Reissner Nordstr\"om (RN) black hole metric, where the charge contributes independently via the term $Q/r^2$, in this case, the charge contribution is coupled with the mass, $M_{ADM}$, and additional parameter $\xi$, as shown in Eq. \eqref{eq:16}. Additionally, the contribution of the third term in Eq. \eqref{eq:15} is negligible because $Q \ll M$, and its impact is proportional to the fourth power of the radial position. However, the effect of the charge on the strong deflection angle (SDA) can still be analyzed by considering only the first two terms of the metric function $A(r)$.

Now, we have an approximate expression of the metric function given as,
\begin{equation}
    A(r) = 1-\frac{2M_{ADM}}{r}, \ \  B(r) = \frac{1}{A(r)},
\end{equation}
where it resembles a Schwarzschild metric. As one of the important properties of the black hole metric, we have to calculate the photon sphere, using the equation $ r_{\textit{ph}}A'( r_{\textit{ph}}) - 2 A( r_{\textit{ph}}) = 0$, we approximate it to be, 
\begin{equation}
    r_{\textit{ph}} = 3M_{ADM}.
\end{equation}
Evaluating the strong deflection constant coefficients $\bar{a}$ and the logarithmic argument of $\bar{b}$ using the equations in \eqref{abar} and \eqref{logarg} yields,
\begin{equation}
    \begin{split}
        \bar{a} &= 1\\
        \bar{b} & =\log\left[6\right]+I_{R}(r_\text{ps})-\pi.
    \end{split}
\end{equation}
The calculated results for $\bar{a}$ and $\bar{b}$ are consistent with the strong Schwarzschild deflection coefficient, as shown in Ref. \cite{Bozza:2002zj}. It implies that the strong lensing of the NLED black hole metric is analogous to the Schwarzschild black hole.

We evaluate $I_{R}$ to complete the deflection angle expression. Since the whole metric is reduced to a Schwarzschild-like metric, where the parameter $\xi$ is embedded in the modified mass, $M_{ADM}$, the result of $I_{R}$ is consistent to V. Bozza's calculation in Ref. \cite{Bozza:2002zj}, hence evaluating the integral $I_{R}(r_{0})$ as $r_{0}\rightarrow r_{ps}$ yields,
\begin{equation}
    I_{R}(r_{\text{ps}}) = 2\log\left(6(2-\sqrt{3}) \right).
\end{equation}
Collecting all the necessary expressions and evaluating the integral, we retrieved the strong deflection angle,
\begin{equation}
\begin{split}
\hat{\alpha}_{\text{str}}
&= -\log\!\Bigl(\frac{b_0}{b_{\rm crit}} - 1\Bigr)
  + \log\!\bigl[216\,(7 - 4\sqrt{3})\bigr]
  - \pi \\[4pt]
&\quad
  + O\!\Bigl(\frac{b_0}{b_{\rm crit}} - 1\Bigr)
    \log\!\Bigl(\frac{b_0}{b_{\rm crit}} - 1\Bigr).
\end{split}
\end{equation}

To calculate the impact parameter, we utilize equation \eqref{bcrit} and note that $r \rightarrow r_{0}$ this will generate the impact parameter for the closest approach,
\begin{equation}
    b_{0}^{2} = \frac{r_{0}^{3}}{r_{0} - 2M_{ADM}},
\end{equation}
when the $r_{0}\rightarrow 3M_{ADM}$ it yields the critical impact parameter, $b_{crit} = 3\sqrt{3}M_{ADM}$.

\begin{figure}[htbp]
	\begin{center}
        {\includegraphics[width=0.48\textwidth]{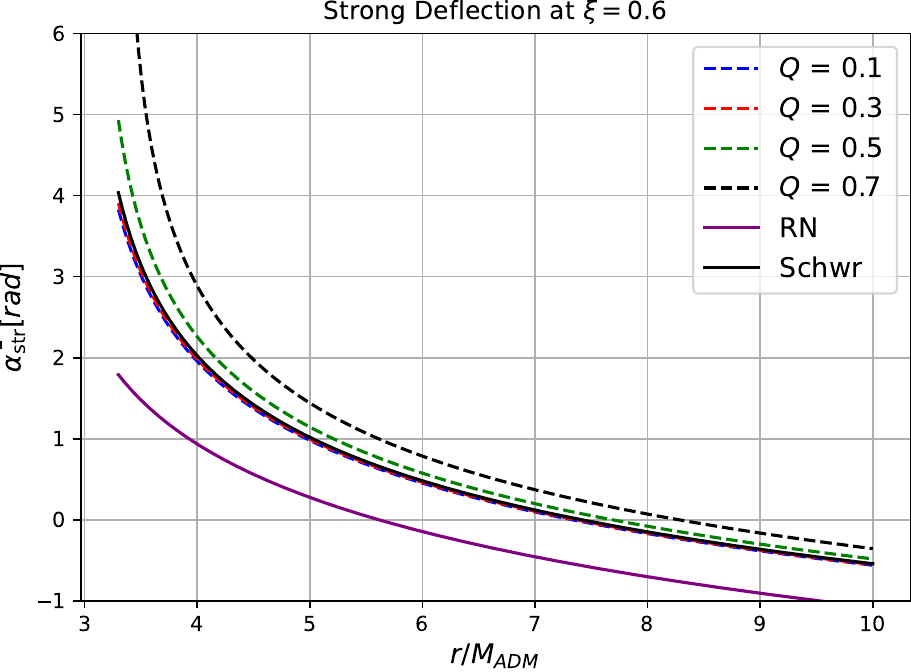}} 
         {\includegraphics[width=0.48\textwidth]{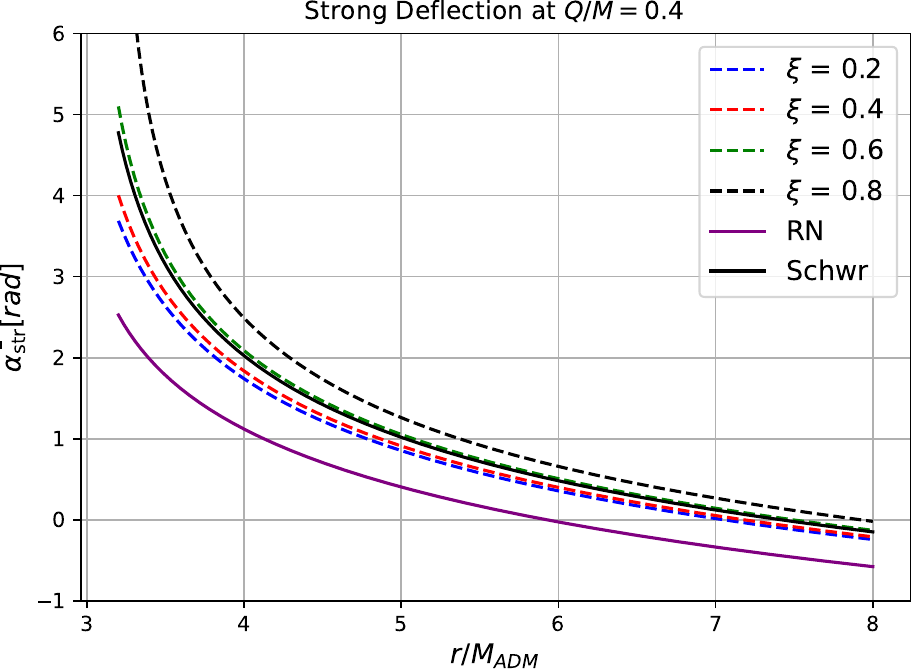}} 
	\end{center}
	\caption{Strong deflection angle with varying parameters $Q$ and $\xi$. In the left figure we use the $\xi = 0.6$, while on the right hand side we use $Q/M = 0.4$}
    \label{fig.SDA1}
\end{figure}

Figure \ref{fig.SDA1} illustrates the influence of the black hole parameters $\xi$ and $Q$ within the strong-field regime. The left panel demonstrates that, at specific values of $\xi$, the charge $Q$ induces a noticeable deviation from the predictions of the Schwarzschild and RN black holes. This observation indicates that the charge $Q$ exerts a significant influence on the photon sphere, a region where the deflection angle diverges. Moreover, the effects of variations in the parameter $\xi$ are found to align consistently with changes in $Q$, highlighting the coupled interplay between these two parameters.

\textcolor{black}{Using the photon ring size, we can predict the possible values of $\xi$. In the case of M87*, a more precise measurement of the photon ring radius improves the accuracy of these constraints. In particular, for the NLED parameter to remain physically meaningful (i.e., real and positive), the photon ring must be larger than $20 \ \mu\text{as}$. Smaller than this, the NLED parameter becomes a complex number. If the photon ring size is constrained more precisely to $20 \ \mu\text{as} < \theta_{\text{p}} < 21 \ \mu\text{as}$, the resulting range for the NLED parameter is $3.8 \times 10^{-21} \leq \xi \leq 4.8 \times 10^{-21}$. For Sgr A*, applying its photon ring constraints gives a range of $1.2 \times 10^{-20} \leq \xi \leq 5.7 \times 10^{-21}$.}}

\section{Shadow Cast} \label{sec5}
\begin{figure}
    \centering
\includegraphics[width=0.5\textwidth]{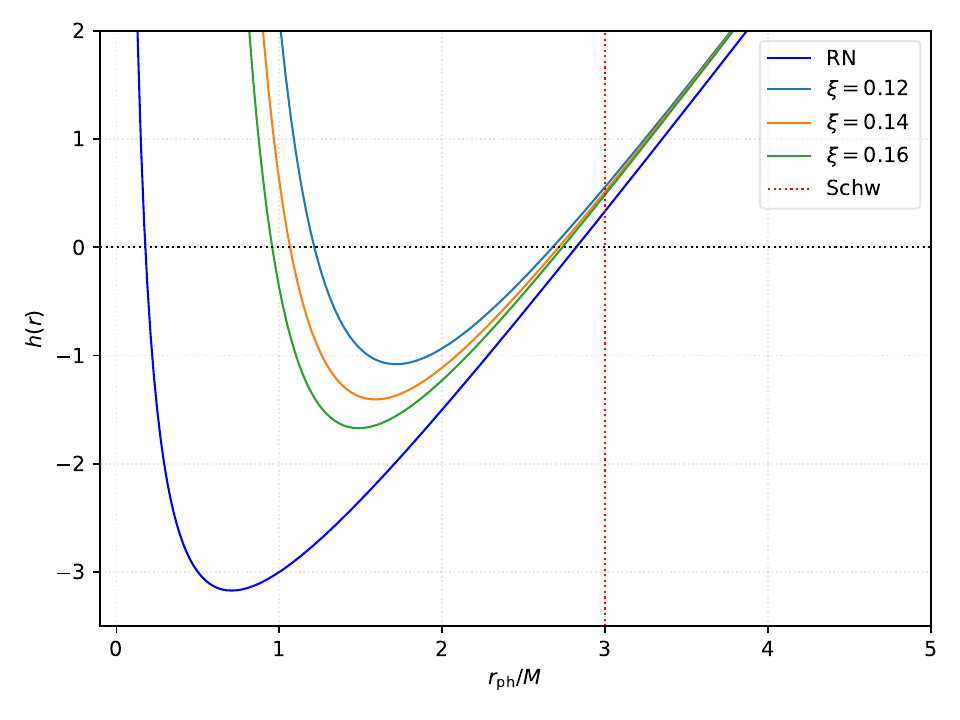}
    \caption{Behavior of the photon sphere. Here, we have chosen an arbitrary value of $Q/M = 0.50$ to enable comparison. The vertical dotted line is the photon sphere radius for the Schwarzschild case.}
    \label{fig:5}
\end{figure}

In studying the shadow of any static and spherically symmetric black hole, methods are widely known \cite{Perlick:2021aok,Perlick:2015vta} and used in many studies present in the literature. We follow such methods in this work and summarize the most important part of the expressions needed to study the black hole shadow. One usually begins with the Lagrangian or Hamiltonian for light rays, then obtains the equations of motion, and the orbit equation. From the orbit equation, the photon sphere can be solved by taking the derivative of the function.
\begin{equation}
    h\left(r\right) = \frac{C\left(r\right)}{A\left(r\right)}, \label{eq:62}
\end{equation}
with respect to $r$, and setting to zero. It results to the expression:
\begin{equation}
    h'\left(r\right) = C'\left(r\right) A\left(r\right) - C\left(r\right)
    A'\left(r\right) = 0.\label{eq:63}
\end{equation}
Depending on the metric functions' complexity, we may or may not obtain some analytical solution to $r_{\rm co}$. The derivation for the shadow becomes simple if there is an analytical solution for  $r_{\rm co}$. From Eqs. (\ref{eq:15}-\ref{eq:16}), it is not possible to obtain an analytical solution. To know the behavior of the photon sphere under the influence of the parameter $\xi$, we rely on numerical considerations shown in Fig. \ref{fig:5}. We found that $\xi$ causes a smaller photon sphere radius as compared to the RN case. The radius is even smaller, as $\xi$ decreases.

\begin{figure}
    \centering
\includegraphics[width=0.5\textwidth]{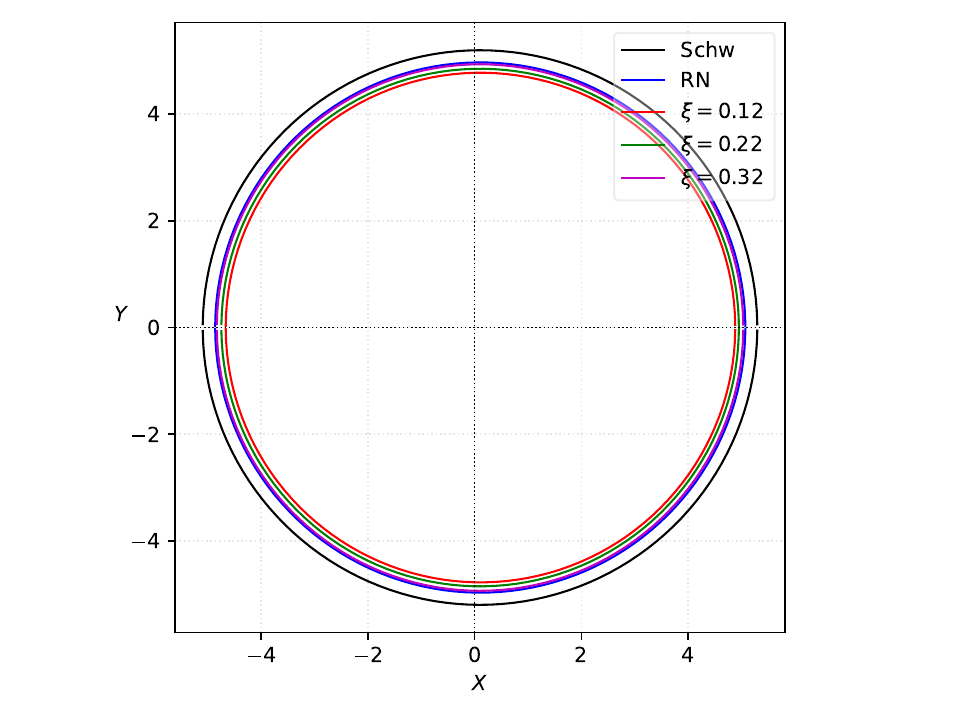}
    \includegraphics[width=0.48\textwidth]{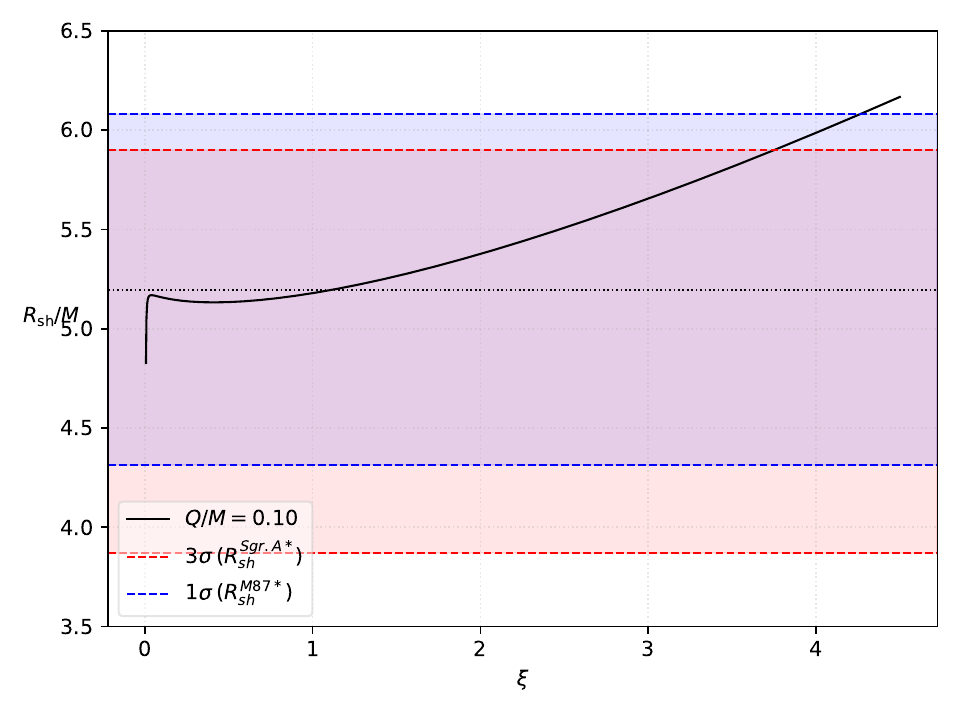}
    \caption{Left: Behavior of the shadow cast. Here, we have chosen an arbitrary value of $Q/M = 0.50$ to enable comparison. The plot shown is in terms of the Cartesian system. Right: EHT constraints for the parameter $\xi$. Here, we assume a small value of charge $Q = 0.10M$, and $r_{\rm obs}\rightarrow \infty$ so that $A(r_{\rm obs}) = 1$.}
    \label{fig:6}
\end{figure}

The most important quantity in the formation of the shadow is the critical impact parameter of light, defined by,
\begin{equation}
\label{bcrit}
    b_{\rm crit}^2 = \frac{C\left(r_{\rm co}\right)}{A\left(r_{\rm co}\right)},
\end{equation}
and the analytic expression of the shadow can then be calculated using:
\begin{equation}
    R_{\rm sh} = b_{\rm crit}\sqrt{A\left(r_{\rm obs}\right)}.
\end{equation}
Again, while the resulting expression can be quite complicated to study analytically, we used a numerical approach to visualize the shadow cast. This is shown in Fig. \ref{fig:6}.
First, we observe that while the deviation from the RN case is quite noticeable in $r_{\rm co}$ for $\xi = 0.12$ to $\xi=0.16$, the deviation in the shadow cast is not too noticeable. Hence, to show strong deviation, the values of $\xi$ are increased on the plot. In essence, we can tell that decreasing the photon sphere results in a decrease in the shadow radius. Furthermore, the shadow radius also decreases as we decrease $\xi$.

Let us now turn our attention to constraining the parameter $\xi$ through the lens of Event Horizon Telescope (EHT) observations. The Schwarzschild shadow radii inferred for Sgr. A* and M87*, within the $2\sigma$ and $1\sigma$ confidence intervals respectively, are bounded as $4.209M \leq R_{\rm Schw} \leq 5.560M$ for Sgr. A* \cite{Vagnozzi:2022moj} and $4.313M \leq R_{\rm Schw} \leq 6.079M$ for M87* \cite{EventHorizonTelescope:2021dqv}. Defining $\Delta$ as the perturbative deviation from the Schwarzschild shadow radius, we obtain the constraints $-0.364 \leq \Delta/M \leq 0.987$ for Sgr. A*, and a symmetric deviation of $\Delta/M = \pm 0.883$ for M87*. These observational bounds will serve as phenomenological filters in narrowing the viable parameter space for $\xi$. We numerically plot the results found in the right panel of Fig. \ref{fig:6}. Since the metric function has no proportional dependence on $r_{\rm obs}$, it is safe to assume the approximation $r_{\rm obs}\rightarrow \infty$ for both astrophysical black holes. As a consequence, we only need one curve to describe how the shadow radius varies with $\xi$. We also approximate a low value for $Q$ to theoretically see its effect on the astrophysical settings. In general, we see that as $\xi$ increases, the shadow size increases, especially at larger values of $\xi$. This is also consistent in the left panel. As a result, we found that the upper bound of $\xi$ for Sgr. A* is $\sim 3.78$, while that in M87* is $\sim 4.29$. These thresholds effectively constrain the degree of nonlinear corrections permissible in realistic black hole models. Below these bounds, the model remains observationally viable. Physically, these results emphasize that while $ \xi $ cannot vanish in this NED black hole framework, its small values are still capable of producing realistic black hole shadows, especially when the magnetic charge $ Q $ is small. In this limit, the deviation from the Reissner–Nordstr\"om case becomes subtle, yet the model retains its distinct NED signature, which could be probed by future high-resolution imaging of black hole environments.

\textcolor{black}{ In particular, the next-generation Event Horizon Telescope (ngEHT), operating at higher frequencies (e.g., 345 GHz), will significantly improve angular resolution, polarization sensitivity, and time-domain imaging capabilities, enabling measurements of black hole shadow features at the few-percent precision level \cite{Blackburn:2019bly,Doeleman:2023kzg,Raymond:2021syf}. Furthermore, proposed space-based VLBI missions such as Millimetron or the Event Horizon Imager could achieve resolutions of a few microarcseconds, dramatically enhancing our ability to detect subtle deviations from the predictions of general relativity due to nonlinear electrodynamics (NED) \cite{Johnson:2019ljv,Kardashev:2014sjq}. Complementary near-infrared observations from GRAVITY+ will provide independent constraints on the spacetime geometry near black holes, further strengthening future tests of NED models \cite{Chael:2022meh}. These advancements collectively promise to either significantly tighten the constraints on NED parameters or reveal compelling signatures of new physics in strong gravitational fields.}

\section{Conclusion} \label{conc}

In this work, we have investigated the gravitational lensing characteristics and shadow formation of a magnetically charged black hole within the framework of nonlinear electrodynamics. Using a geometric approach based on the Gauss-Bonnet theorem, we derived the trajectory equation, the Gaussian curvature, and an analytical expression for the weak-field deflection angle of light propagating in the equatorial plane. Our analysis shows that the standard Schwarzschild deflection term, \(4M/b\), is augmented by additional corrections that depend on both the magnetic charge \(Q\) and the nonlinear coupling parameter \(\xi\). These extra contributions, which appear as power-law and logarithmic terms, highlight the nontrivial influence of nonlinear electromagnetic effects on the spacetime geometry and emphasize modifications in the Reissner--Nordstr\"om (RN) metric relative to the Schwarzschild case.

In the strong-field regime, we comprehensively analyzed light ray behavior near the photon sphere using the strong deflection limit formalism. In this regime, the deflection angle exhibits the characteristic logarithmic divergence as the impact parameter approaches its critical value. The derived strong deflection coefficients, $\bar{a}$ and $\bar{b}$, closely resemble those in the Schwarzschild strong deflection equation. However, the presence of a nonlinear electrodynamic (NED) charge introduces a significant deviation, enhancing the bending of light compared to the standard Reissner--Nordstr\"om (RN) case. This enhancement, arising from the interplay between the charge $Q$ and the NLED parameter $\xi$, not only modifies the near-horizon geodesics but also has profound implications for the structure of the photon sphere and the resulting black hole shadow.

\textcolor{black}{Furthermore, the NLED parameter $\xi$ plays an important role in refining the precision of constraints for the M87$^{*}$ black hole. Since $\xi$ must remain real and positive, its lower bound is necessarily shifted upward, exceeding the value measured by the Event Horizon Telescope (EHT). Notably, the inferred values of $\xi$ for both M87$^{*}$ and Sgr A$^{*}$ consistently fall within the $10^{-21}$ order of magnitude, reinforcing the robustness of the theoretical framework and its alignment with observational data.}

Our numerical analysis indicates that increasing the coupling parameter \(\xi\) or the charge \(Q\) leads to a reduction in the radius of the photon sphere, thereby producing a smaller shadow. Such behavior distinguishes nonlinear electrodynamic black holes from their classical counterparts and offers a potential observational signature that could be tested in future high-resolution imaging experiments. Moreover, the tight constraints imposed on \(\xi\) by solar system light-bending observations ensure that any deviations from classical predictions remain within acceptable observational limits, thereby reinforcing the consistency of our approach.

The results presented here provide some hints about the key importance of the modifications induced by nonlinear electrodynamics on both the gravitational lensing and shadow characteristics of black holes. They demonstrate the potential of precision lensing and shadow observations to probe deviations from standard general relativity and to test the viability of alternative theories of gravity in extreme astrophysical environments.

\textcolor{black}{Upcoming enhancements from next-generation VLBI facilities, such as the next-generation Event Horizon Telescope (ngEHT), GRAVITY+, and space-based VLBI missions, promise significantly improved angular resolution, polarization sensitivity, and multi-frequency imaging. These developments are expected to sharpen shadow measurements to the level of a few microarcseconds, enabling precision tests of deviations from general relativity induced by nonlinear electrodynamics (NED). Consequently, future observational data could considerably tighten constraints on the NED coupling parameter $\xi$, or even provide the first definitive detection of NED signatures in black hole shadows, representing a significant advance in our understanding of strong field gravity.}

\acknowledgements
R. P. and A. \"O. would like to acknowledge networking support of the COST Action CA21106 - COSMIC WISPers in the Dark Universe: Theory, astrophysics and experiments (CosmicWISPers), the COST Action CA22113 - Fundamental challenges in theoretical physics (THEORY-CHALLENGES), the COST Action CA21136 - Addressing observational tensions in cosmology with systematics and fundamental physics (CosmoVerse), the COST Action CA23130 - Bridging high and low energies in search of quantum gravity (BridgeQG), and the COST Action CA23115 - Relativistic Quantum Information (RQI) funded by COST (European Cooperation in Science and Technology). A. \"O. also thanks to EMU, TUBITAK, ULAKBIM (Turkiye) and SCOAP3 (Switzerland) for their support.

\bibliography{references}

\begin{thebibliography}{120}%
\makeatletter
\providecommand \@ifxundefined [1]{%
 \@ifx{#1\undefined}
}%
\providecommand \@ifnum [1]{%
 \ifnum #1\expandafter \@firstoftwo
 \else \expandafter \@secondoftwo
 \fi
}%
\providecommand \@ifx [1]{%
 \ifx #1\expandafter \@firstoftwo
 \else \expandafter \@secondoftwo
 \fi
}%
\providecommand \natexlab [1]{#1}%
\providecommand \enquote  [1]{``#1''}%
\providecommand \bibnamefont  [1]{#1}%
\providecommand \bibfnamefont [1]{#1}%
\providecommand \citenamefont [1]{#1}%
\providecommand \href@noop [0]{\@secondoftwo}%
\providecommand \href [0]{\begingroup \@sanitize@url \@href}%
\providecommand \@href[1]{\@@startlink{#1}\@@href}%
\providecommand \@@href[1]{\endgroup#1\@@endlink}%
\providecommand \@sanitize@url [0]{\catcode `\\12\catcode `\$12\catcode `\&12\catcode `\#12\catcode `\^12\catcode `\_12\catcode `\%12\relax}%
\providecommand \@@startlink[1]{}%
\providecommand \@@endlink[0]{}%
\providecommand \url  [0]{\begingroup\@sanitize@url \@url }%
\providecommand \@url [1]{\endgroup\@href {#1}{\urlprefix }}%
\providecommand \urlprefix  [0]{URL }%
\providecommand \Eprint [0]{\href }%
\providecommand \doibase [0]{http://dx.doi.org/}%
\providecommand \selectlanguage [0]{\@gobble}%
\providecommand \bibinfo  [0]{\@secondoftwo}%
\providecommand \bibfield  [0]{\@secondoftwo}%
\providecommand \translation [1]{[#1]}%
\providecommand \BibitemOpen [0]{}%
\providecommand \bibitemStop [0]{}%
\providecommand \bibitemNoStop [0]{.\EOS\space}%
\providecommand \EOS [0]{\spacefactor3000\relax}%
\providecommand \BibitemShut  [1]{\csname bibitem#1\endcsname}%
\let\auto@bib@innerbib\@empty
\bibitem [{\citenamefont {Kline}(1972)}]{klein}%
  \BibitemOpen
  \bibfield  {author} {\bibinfo {author} {\bibfnamefont {M.}~\bibnamefont {Kline}},\ }\href@noop {} {\emph {\bibinfo {title} {Mathematical Thought from Ancient to Modern Times}}}\ (\bibinfo  {publisher} {Oxford University Press},\ \bibinfo {address} {Oxford},\ \bibinfo {year} {1972})\BibitemShut {NoStop}%
\bibitem [{\citenamefont {Bandyopadhyay}\ \emph {et~al.}(2023)\citenamefont {Bandyopadhyay}, \citenamefont {Dacorogna}, \citenamefont {Matveev},\ and\ \citenamefont {Troyanov}}]{Reiman}%
  \BibitemOpen
  \bibfield  {author} {\bibinfo {author} {\bibfnamefont {S.}~\bibnamefont {Bandyopadhyay}}, \bibinfo {author} {\bibfnamefont {B.}~\bibnamefont {Dacorogna}}, \bibinfo {author} {\bibfnamefont {V.S.}\ \bibnamefont {Matveev}}, \ and\ \bibinfo {author} {\bibfnamefont {M.}~\bibnamefont {Troyanov}},\ }\bibfield  {title} {\enquote {\bibinfo {title} {Bernhard riemann 1861 revisited: existence of flat coordinates for an arbitrary bilinear form},}\ }\href {\doibase 10.1007/s00209-023-03335-1} {\bibfield  {journal} {\bibinfo  {journal} {Mathematische Zeitschrift}\ }\textbf {\bibinfo {volume} {305}} (\bibinfo {year} {2023}),\ 10.1007/s00209-023-03335-1},\ \Eprint {http://arxiv.org/abs/arXiv:2109.03098 [math.DG]} {arXiv:2109.03098 [math.DG]} \BibitemShut {NoStop}%
\bibitem [{\citenamefont {Einstein}(1916)}]{A.Einstein}%
  \BibitemOpen
  \bibfield  {author} {\bibinfo {author} {\bibfnamefont {Albert}\ \bibnamefont {Einstein}},\ }\bibfield  {title} {\enquote {\bibinfo {title} {{The foundation of the general theory of relativity.}}}\ }\href {\doibase 10.1002/andp.19163540702} {\bibfield  {journal} {\bibinfo  {journal} {Annalen Phys.}\ }\textbf {\bibinfo {volume} {49}},\ \bibinfo {pages} {769--822} (\bibinfo {year} {1916})}\BibitemShut {NoStop}%
\bibitem [{\citenamefont {Misner}\ \emph {et~al.}(1973)\citenamefont {Misner}, \citenamefont {Thorne},\ and\ \citenamefont {Wheeler}}]{Thorne}%
  \BibitemOpen
  \bibfield  {author} {\bibinfo {author} {\bibfnamefont {C.~W.}\ \bibnamefont {Misner}}, \bibinfo {author} {\bibfnamefont {K.~S.}\ \bibnamefont {Thorne}}, \ and\ \bibinfo {author} {\bibfnamefont {J.~A.}\ \bibnamefont {Wheeler}},\ }\href@noop {} {\emph {\bibinfo {title} {Gravitation}}}\ (\bibinfo  {publisher} {Freeman W. H. and Company},\ \bibinfo {address} {San Francisco},\ \bibinfo {year} {1973})\BibitemShut {NoStop}%
\bibitem [{\citenamefont {Rindler}(1977)}]{Rindler}%
  \BibitemOpen
  \bibfield  {author} {\bibinfo {author} {\bibfnamefont {W.}~\bibnamefont {Rindler}},\ }\href@noop {} {\emph {\bibinfo {title} {Essential Relativity: Special, General, and Cosmological}}},\ \bibinfo {edition} {2nd}\ ed.\ (\bibinfo  {publisher} {Springer Verlag},\ \bibinfo {year} {1977})\BibitemShut {NoStop}%
\bibitem [{\citenamefont {Dyson}\ \emph {et~al.}(1920)\citenamefont {Dyson}, \citenamefont {Eddington},\ and\ \citenamefont {Davidson}}]{Dyson}%
  \BibitemOpen
  \bibfield  {author} {\bibinfo {author} {\bibfnamefont {F.~W.}\ \bibnamefont {Dyson}}, \bibinfo {author} {\bibfnamefont {A.~S.}\ \bibnamefont {Eddington}}, \ and\ \bibinfo {author} {\bibfnamefont {C.}~\bibnamefont {Davidson}},\ }\bibfield  {title} {\enquote {\bibinfo {title} {A determination of the deflection of light by the sun's gravitational field, from observations made at the total eclipse of may 29, 1919},}\ }\href@noop {} {\bibfield  {journal} {\bibinfo  {journal} {Philosophical Transactions of the Royal Society of London. Series A, Containing Papers of a Mathematical or Physical Character}\ }\textbf {\bibinfo {volume} {220}},\ \bibinfo {pages} {291--333} (\bibinfo {year} {1920})}\BibitemShut {NoStop}%
\bibitem [{\citenamefont {Tsukamoto}\ \emph {et~al.}(2012)\citenamefont {Tsukamoto}, \citenamefont {Harada},\ and\ \citenamefont {Yajima}}]{Harada}%
  \BibitemOpen
  \bibfield  {author} {\bibinfo {author} {\bibfnamefont {Naoki}\ \bibnamefont {Tsukamoto}}, \bibinfo {author} {\bibfnamefont {Tomohiro}\ \bibnamefont {Harada}}, \ and\ \bibinfo {author} {\bibfnamefont {Kohji}\ \bibnamefont {Yajima}},\ }\bibfield  {title} {\enquote {\bibinfo {title} {{Can we distinguish between black holes and wormholes by their Einstein ring systems?}}}\ }\href {\doibase 10.1103/PhysRevD.86.104062} {\bibfield  {journal} {\bibinfo  {journal} {Phys. Rev. D}\ }\textbf {\bibinfo {volume} {86}},\ \bibinfo {pages} {104062} (\bibinfo {year} {2012})},\ \Eprint {http://arxiv.org/abs/1207.0047} {arXiv:1207.0047 [gr-qc]} \BibitemShut {NoStop}%
\bibitem [{\citenamefont {Keeton}\ \emph {et~al.}(1998)\citenamefont {Keeton}, \citenamefont {Kochanek},\ and\ \citenamefont {Falco}}]{keeton}%
  \BibitemOpen
  \bibfield  {author} {\bibinfo {author} {\bibfnamefont {C.~R.}\ \bibnamefont {Keeton}}, \bibinfo {author} {\bibfnamefont {C.~S.}\ \bibnamefont {Kochanek}}, \ and\ \bibinfo {author} {\bibfnamefont {E.~E.}\ \bibnamefont {Falco}},\ }\bibfield  {title} {\enquote {\bibinfo {title} {{The Optical properties of gravitational lens galaxies as a probe of galaxy structure and evolution}},}\ }\href {\doibase 10.1086/306502} {\bibfield  {journal} {\bibinfo  {journal} {Astrophys. J.}\ }\textbf {\bibinfo {volume} {509}},\ \bibinfo {pages} {561--578} (\bibinfo {year} {1998})},\ \Eprint {http://arxiv.org/abs/astro-ph/9708161} {arXiv:astro-ph/9708161} \BibitemShut {NoStop}%
\bibitem [{\citenamefont {Virbhadra}\ and\ \citenamefont {Ellis}(2000)}]{Ellis}%
  \BibitemOpen
  \bibfield  {author} {\bibinfo {author} {\bibfnamefont {K.~S.}\ \bibnamefont {Virbhadra}}\ and\ \bibinfo {author} {\bibfnamefont {George F.~R.}\ \bibnamefont {Ellis}},\ }\bibfield  {title} {\enquote {\bibinfo {title} {{Schwarzschild black hole lensing}},}\ }\href {\doibase 10.1103/PhysRevD.62.084003} {\bibfield  {journal} {\bibinfo  {journal} {Phys. Rev. D}\ }\textbf {\bibinfo {volume} {62}},\ \bibinfo {pages} {084003} (\bibinfo {year} {2000})},\ \Eprint {http://arxiv.org/abs/astro-ph/9904193} {arXiv:astro-ph/9904193} \BibitemShut {NoStop}%
\bibitem [{\citenamefont {Khodabakhshi}\ and\ \citenamefont {Mann}(2021)}]{keeton1}%
  \BibitemOpen
  \bibfield  {author} {\bibinfo {author} {\bibfnamefont {H.}~\bibnamefont {Khodabakhshi}}\ and\ \bibinfo {author} {\bibfnamefont {Robert~B.}\ \bibnamefont {Mann}},\ }\bibfield  {title} {\enquote {\bibinfo {title} {{Gravitational Lensing by Black Holes in Einstein Quartic Gravity}},}\ }\href {\doibase 10.1103/PhysRevD.103.024017} {\bibfield  {journal} {\bibinfo  {journal} {Phys. Rev. D}\ }\textbf {\bibinfo {volume} {103}},\ \bibinfo {pages} {024017} (\bibinfo {year} {2021})},\ \Eprint {http://arxiv.org/abs/2007.05341} {arXiv:2007.05341 [gr-qc]} \BibitemShut {NoStop}%
\bibitem [{\citenamefont {Virbhadra}\ and\ \citenamefont {Keeton}(2008)}]{keeton2}%
  \BibitemOpen
  \bibfield  {author} {\bibinfo {author} {\bibfnamefont {K.~S.}\ \bibnamefont {Virbhadra}}\ and\ \bibinfo {author} {\bibfnamefont {C.~R.}\ \bibnamefont {Keeton}},\ }\bibfield  {title} {\enquote {\bibinfo {title} {{Time delay and magnification centroid due to gravitational lensing by black holes and naked singularities}},}\ }\href {\doibase 10.1103/PhysRevD.77.124014} {\bibfield  {journal} {\bibinfo  {journal} {Phys. Rev. D}\ }\textbf {\bibinfo {volume} {77}},\ \bibinfo {pages} {124014} (\bibinfo {year} {2008})},\ \Eprint {http://arxiv.org/abs/0710.2333} {arXiv:0710.2333 [gr-qc]} \BibitemShut {NoStop}%
\bibitem [{\citenamefont {Chen}\ \emph {et~al.}(2009)\citenamefont {Chen}, \citenamefont {Cai}, \citenamefont {Ko}, \citenamefont {Li}, \citenamefont {Shen},\ and\ \citenamefont {Xu}}]{chen}%
  \BibitemOpen
  \bibfield  {author} {\bibinfo {author} {\bibfnamefont {Lie-Wen}\ \bibnamefont {Chen}}, \bibinfo {author} {\bibfnamefont {Bao-Jun}\ \bibnamefont {Cai}}, \bibinfo {author} {\bibfnamefont {Che~Ming}\ \bibnamefont {Ko}}, \bibinfo {author} {\bibfnamefont {Bao-An}\ \bibnamefont {Li}}, \bibinfo {author} {\bibfnamefont {Chun}\ \bibnamefont {Shen}}, \ and\ \bibinfo {author} {\bibfnamefont {Jun}\ \bibnamefont {Xu}},\ }\bibfield  {title} {\enquote {\bibinfo {title} {Higher-order effects on the incompressibility of isospin asymmetric nuclear matter},}\ }\href {\doibase 10.1103/PhysRevC.80.014322} {\bibfield  {journal} {\bibinfo  {journal} {Phys. Rev. C}\ }\textbf {\bibinfo {volume} {80}},\ \bibinfo {pages} {014322} (\bibinfo {year} {2009})}\BibitemShut {NoStop}%
\bibitem [{\citenamefont {Sharif}\ and\ \citenamefont {Yousaf}(2015)}]{sharif}%
  \BibitemOpen
  \bibfield  {author} {\bibinfo {author} {\bibfnamefont {M.}~\bibnamefont {Sharif}}\ and\ \bibinfo {author} {\bibfnamefont {Z.}~\bibnamefont {Yousaf}},\ }\bibfield  {title} {\enquote {\bibinfo {title} {{Radiating cylindrical gravitational collapse with structure scalars in f(R) gravity}},}\ }\href {\doibase 10.1007/s10509-015-2270-2} {\bibfield  {journal} {\bibinfo  {journal} {Astrophys. Space Sci.}\ }\textbf {\bibinfo {volume} {357}},\ \bibinfo {pages} {49} (\bibinfo {year} {2015})}\BibitemShut {NoStop}%
\bibitem [{\citenamefont {Cao}\ and\ \citenamefont {Xie}(2018)}]{Cao}%
  \BibitemOpen
  \bibfield  {author} {\bibinfo {author} {\bibfnamefont {Wei-Guang}\ \bibnamefont {Cao}}\ and\ \bibinfo {author} {\bibfnamefont {Yi}~\bibnamefont {Xie}},\ }\bibfield  {title} {\enquote {\bibinfo {title} {{Weak deflection gravitational lensing for photons coupled to Weyl tensor in a Schwarzschild black hole}},}\ }\href {\doibase 10.1140/epjc/s10052-018-5684-5} {\bibfield  {journal} {\bibinfo  {journal} {Eur. Phys. J. C}\ }\textbf {\bibinfo {volume} {78}},\ \bibinfo {pages} {191} (\bibinfo {year} {2018})}\BibitemShut {NoStop}%
\bibitem [{\citenamefont {Bisnovatyi-Kogan}\ and\ \citenamefont {Tsupko}(2017)}]{kogan}%
  \BibitemOpen
  \bibfield  {author} {\bibinfo {author} {\bibfnamefont {Gennady~S.}\ \bibnamefont {Bisnovatyi-Kogan}}\ and\ \bibinfo {author} {\bibfnamefont {Oleg~Yu.}\ \bibnamefont {Tsupko}},\ }\bibfield  {title} {\enquote {\bibinfo {title} {Gravitational lensing in presence of plasma: Strong lens systems, black hole lensing and shadow},}\ }\href {\doibase 10.3390/universe3030057} {\bibfield  {journal} {\bibinfo  {journal} {Universe}\ }\textbf {\bibinfo {volume} {3}} (\bibinfo {year} {2017}),\ 10.3390/universe3030057}\BibitemShut {NoStop}%
\bibitem [{\citenamefont {Bisnovatyi-Kogan}\ and\ \citenamefont {Tsupko}(2010)}]{roy}%
  \BibitemOpen
  \bibfield  {author} {\bibinfo {author} {\bibfnamefont {G.~S.}\ \bibnamefont {Bisnovatyi-Kogan}}\ and\ \bibinfo {author} {\bibfnamefont {O.~Yu.}\ \bibnamefont {Tsupko}},\ }\bibfield  {title} {\enquote {\bibinfo {title} {{Gravitational lensing in a non-uniform plasma}},}\ }\href {\doibase 10.1111/j.1365-2966.2010.16290.x} {\bibfield  {journal} {\bibinfo  {journal} {Mon. Not. Roy. Astron. Soc.}\ }\textbf {\bibinfo {volume} {404}},\ \bibinfo {pages} {1790--1800} (\bibinfo {year} {2010})},\ \Eprint {http://arxiv.org/abs/1006.2321} {arXiv:1006.2321 [astro-ph.CO]} \BibitemShut {NoStop}%
\bibitem [{\citenamefont {Refsdal}(1964)}]{refsdal}%
  \BibitemOpen
  \bibfield  {author} {\bibinfo {author} {\bibfnamefont {S.}~\bibnamefont {Refsdal}},\ }\bibfield  {title} {\enquote {\bibinfo {title} {{On the possibility of determining Hubble's parameter and the masses of galaxies from the gravitational lens effect}},}\ }\href@noop {} {\bibfield  {journal} {\bibinfo  {journal} {Mon. Not. Roy. Astron. Soc.}\ }\textbf {\bibinfo {volume} {128}},\ \bibinfo {pages} {307} (\bibinfo {year} {1964})}\BibitemShut {NoStop}%
\bibitem [{\citenamefont {Bartelmann}\ and\ \citenamefont {Schneider}(2001)}]{Bartelmann:1999yn}%
  \BibitemOpen
  \bibfield  {author} {\bibinfo {author} {\bibfnamefont {Matthias}\ \bibnamefont {Bartelmann}}\ and\ \bibinfo {author} {\bibfnamefont {Peter}\ \bibnamefont {Schneider}},\ }\bibfield  {title} {\enquote {\bibinfo {title} {{Weak gravitational lensing}},}\ }\href {\doibase 10.1016/S0370-1573(00)00082-X} {\bibfield  {journal} {\bibinfo  {journal} {Phys. Rept.}\ }\textbf {\bibinfo {volume} {340}},\ \bibinfo {pages} {291--472} (\bibinfo {year} {2001})},\ \Eprint {http://arxiv.org/abs/astro-ph/9912508} {arXiv:astro-ph/9912508} \BibitemShut {NoStop}%
\bibitem [{\citenamefont {Schneider}\ \emph {et~al.}(1992)\citenamefont {Schneider}, \citenamefont {Ehlers},\ and\ \citenamefont {Falco}}]{falco}%
  \BibitemOpen
  \bibfield  {author} {\bibinfo {author} {\bibfnamefont {Peter}\ \bibnamefont {Schneider}}, \bibinfo {author} {\bibfnamefont {J{\"u}rgen}\ \bibnamefont {Ehlers}}, \ and\ \bibinfo {author} {\bibfnamefont {Emilio~E}\ \bibnamefont {Falco}},\ }\href@noop {} {\emph {\bibinfo {title} {Gravitational lenses as astrophysical tools}}}\ (\bibinfo  {publisher} {Springer},\ \bibinfo {year} {1992})\BibitemShut {NoStop}%
\bibitem [{\citenamefont {Bloomer}(2011)}]{Bloomer:2011rd}%
  \BibitemOpen
  \bibfield  {author} {\bibinfo {author} {\bibfnamefont {Cameron}\ \bibnamefont {Bloomer}},\ }\href@noop {} {\enquote {\bibinfo {title} {{Optical Geometry of the Kerr Space-time}},}\ } (\bibinfo {year} {2011}),\ \Eprint {http://arxiv.org/abs/1111.4998} {arXiv:1111.4998 [math-ph]} \BibitemShut {NoStop}%
\bibitem [{\citenamefont {Werner}(2012)}]{werner}%
  \BibitemOpen
  \bibfield  {author} {\bibinfo {author} {\bibfnamefont {M.~C.}\ \bibnamefont {Werner}},\ }\bibfield  {title} {\enquote {\bibinfo {title} {{Gravitational lensing in the Kerr-Randers optical geometry}},}\ }\href {\doibase 10.1007/s10714-012-1458-9} {\bibfield  {journal} {\bibinfo  {journal} {Gen. Rel. Grav.}\ }\textbf {\bibinfo {volume} {44}},\ \bibinfo {pages} {3047--3057} (\bibinfo {year} {2012})},\ \Eprint {http://arxiv.org/abs/1205.3876} {arXiv:1205.3876 [gr-qc]} \BibitemShut {NoStop}%
\bibitem [{\citenamefont {Godunov}\ \emph {et~al.}(2016)\citenamefont {Godunov}, \citenamefont {Rozanov}, \citenamefont {Vysotsky},\ and\ \citenamefont {Zhemchugov}}]{Godunov:2015nea}%
  \BibitemOpen
  \bibfield  {author} {\bibinfo {author} {\bibfnamefont {S.~I.}\ \bibnamefont {Godunov}}, \bibinfo {author} {\bibfnamefont {A.~N.}\ \bibnamefont {Rozanov}}, \bibinfo {author} {\bibfnamefont {M.~I.}\ \bibnamefont {Vysotsky}}, \ and\ \bibinfo {author} {\bibfnamefont {E.~V.}\ \bibnamefont {Zhemchugov}},\ }\bibfield  {title} {\enquote {\bibinfo {title} {{Extending the Higgs sector: an extra singlet}},}\ }\href {\doibase 10.1140/epjc/s10052-015-3826-6} {\bibfield  {journal} {\bibinfo  {journal} {Eur. Phys. J. C}\ }\textbf {\bibinfo {volume} {76}},\ \bibinfo {pages} {1} (\bibinfo {year} {2016})},\ \Eprint {http://arxiv.org/abs/1503.01618} {arXiv:1503.01618 [hep-ph]} \BibitemShut {NoStop}%
\bibitem [{\citenamefont {Ishihara}\ \emph {et~al.}(2016)\citenamefont {Ishihara}, \citenamefont {Suzuki}, \citenamefont {Ono}, \citenamefont {Kitamura},\ and\ \citenamefont {Asada}}]{Ishihara}%
  \BibitemOpen
  \bibfield  {author} {\bibinfo {author} {\bibfnamefont {Asahi}\ \bibnamefont {Ishihara}}, \bibinfo {author} {\bibfnamefont {Yusuke}\ \bibnamefont {Suzuki}}, \bibinfo {author} {\bibfnamefont {Toshiaki}\ \bibnamefont {Ono}}, \bibinfo {author} {\bibfnamefont {Takao}\ \bibnamefont {Kitamura}}, \ and\ \bibinfo {author} {\bibfnamefont {Hideki}\ \bibnamefont {Asada}},\ }\bibfield  {title} {\enquote {\bibinfo {title} {{Gravitational bending angle of light for finite distance and the Gauss-Bonnet theorem}},}\ }\href {\doibase 10.1103/PhysRevD.94.084015} {\bibfield  {journal} {\bibinfo  {journal} {Phys. Rev. D}\ }\textbf {\bibinfo {volume} {94}},\ \bibinfo {pages} {084015} (\bibinfo {year} {2016})},\ \Eprint {http://arxiv.org/abs/1604.08308} {arXiv:1604.08308 [gr-qc]} \BibitemShut {NoStop}%
\bibitem [{\citenamefont {Ishihara}\ \emph {et~al.}(2017)\citenamefont {Ishihara}, \citenamefont {Suzuki}, \citenamefont {Ono},\ and\ \citenamefont {Asada}}]{Ishihara:2016sfv}%
  \BibitemOpen
  \bibfield  {author} {\bibinfo {author} {\bibfnamefont {Asahi}\ \bibnamefont {Ishihara}}, \bibinfo {author} {\bibfnamefont {Yusuke}\ \bibnamefont {Suzuki}}, \bibinfo {author} {\bibfnamefont {Toshiaki}\ \bibnamefont {Ono}}, \ and\ \bibinfo {author} {\bibfnamefont {Hideki}\ \bibnamefont {Asada}},\ }\bibfield  {title} {\enquote {\bibinfo {title} {{Finite-distance corrections to the gravitational bending angle of light in the strong deflection limit}},}\ }\href {\doibase 10.1103/PhysRevD.95.044017} {\bibfield  {journal} {\bibinfo  {journal} {Phys. Rev. D}\ }\textbf {\bibinfo {volume} {95}},\ \bibinfo {pages} {044017} (\bibinfo {year} {2017})},\ \Eprint {http://arxiv.org/abs/1612.04044} {arXiv:1612.04044 [gr-qc]} \BibitemShut {NoStop}%
\bibitem [{\citenamefont {Jusufi}\ \emph {et~al.}(2017)\citenamefont {Jusufi}, \citenamefont {Werner}, \citenamefont {Banerjee},\ and\ \citenamefont {\"Ovg\"un}}]{a.ovgun}%
  \BibitemOpen
  \bibfield  {author} {\bibinfo {author} {\bibfnamefont {Kimet}\ \bibnamefont {Jusufi}}, \bibinfo {author} {\bibfnamefont {Marcus~C.}\ \bibnamefont {Werner}}, \bibinfo {author} {\bibfnamefont {Ayan}\ \bibnamefont {Banerjee}}, \ and\ \bibinfo {author} {\bibfnamefont {Ali}\ \bibnamefont {\"Ovg\"un}},\ }\bibfield  {title} {\enquote {\bibinfo {title} {{Light Deflection by a Rotating Global Monopole Spacetime}},}\ }\href {\doibase 10.1103/PhysRevD.95.104012} {\bibfield  {journal} {\bibinfo  {journal} {Phys. Rev. D}\ }\textbf {\bibinfo {volume} {95}},\ \bibinfo {pages} {104012} (\bibinfo {year} {2017})},\ \Eprint {http://arxiv.org/abs/1702.05600} {arXiv:1702.05600 [gr-qc]} \BibitemShut {NoStop}%
\bibitem [{\citenamefont {Goulart}(2018)}]{Goulart:2017iko}%
  \BibitemOpen
  \bibfield  {author} {\bibinfo {author} {\bibfnamefont {Prieslei}\ \bibnamefont {Goulart}},\ }\bibfield  {title} {\enquote {\bibinfo {title} {{Phantom wormholes in Einstein\textendash{}Maxwell-dilaton theory}},}\ }\href {\doibase 10.1088/1361-6382/aa9dfc} {\bibfield  {journal} {\bibinfo  {journal} {Class. Quant. Grav.}\ }\textbf {\bibinfo {volume} {35}},\ \bibinfo {pages} {025012} (\bibinfo {year} {2018})},\ \Eprint {http://arxiv.org/abs/1708.00935} {arXiv:1708.00935 [gr-qc]} \BibitemShut {NoStop}%
\bibitem [{\citenamefont {Ono}\ \emph {et~al.}(2017)\citenamefont {Ono}, \citenamefont {Ishihara},\ and\ \citenamefont {Asada}}]{Ono:2017pie}%
  \BibitemOpen
  \bibfield  {author} {\bibinfo {author} {\bibfnamefont {Toshiaki}\ \bibnamefont {Ono}}, \bibinfo {author} {\bibfnamefont {Asahi}\ \bibnamefont {Ishihara}}, \ and\ \bibinfo {author} {\bibfnamefont {Hideki}\ \bibnamefont {Asada}},\ }\bibfield  {title} {\enquote {\bibinfo {title} {{Gravitomagnetic bending angle of light with finite-distance corrections in stationary axisymmetric spacetimes}},}\ }\href {\doibase 10.1103/PhysRevD.96.104037} {\bibfield  {journal} {\bibinfo  {journal} {Phys. Rev. D}\ }\textbf {\bibinfo {volume} {96}},\ \bibinfo {pages} {104037} (\bibinfo {year} {2017})},\ \Eprint {http://arxiv.org/abs/1704.05615} {arXiv:1704.05615 [gr-qc]} \BibitemShut {NoStop}%
\bibitem [{\citenamefont {Fleischer}\ \emph {et~al.}(2018)\citenamefont {Fleischer}, \citenamefont {Espinosa}, \citenamefont {Jaarsma},\ and\ \citenamefont {Tetlalmatzi-Xolocotzi}}]{Fleischer:2017yox}%
  \BibitemOpen
  \bibfield  {author} {\bibinfo {author} {\bibfnamefont {Robert}\ \bibnamefont {Fleischer}}, \bibinfo {author} {\bibfnamefont {Daniela~Gal\'arraga}\ \bibnamefont {Espinosa}}, \bibinfo {author} {\bibfnamefont {Ruben}\ \bibnamefont {Jaarsma}}, \ and\ \bibinfo {author} {\bibfnamefont {Gilberto}\ \bibnamefont {Tetlalmatzi-Xolocotzi}},\ }\bibfield  {title} {\enquote {\bibinfo {title} {{CP Violation in Leptonic Rare $B^0_s$ Decays as a Probe of New Physics}},}\ }\href {\doibase 10.1140/epjc/s10052-017-5488-z} {\bibfield  {journal} {\bibinfo  {journal} {Eur. Phys. J. C}\ }\textbf {\bibinfo {volume} {78}},\ \bibinfo {pages} {1} (\bibinfo {year} {2018})},\ \Eprint {http://arxiv.org/abs/1709.04735} {arXiv:1709.04735 [hep-ph]} \BibitemShut {NoStop}%
\bibitem [{\citenamefont {Arakida}(2018{\natexlab{a}})}]{arakida}%
  \BibitemOpen
  \bibfield  {author} {\bibinfo {author} {\bibfnamefont {H.}~\bibnamefont {Arakida}},\ }\bibfield  {title} {\enquote {\bibinfo {title} {A note on the definitions of the relativistic perihelion advance},}\ }\href@noop {} {\bibfield  {journal} {\bibinfo  {journal} {General Relativity and Gravitation}\ }\textbf {\bibinfo {volume} {50}},\ \bibinfo {pages} {1} (\bibinfo {year} {2018}{\natexlab{a}})}\BibitemShut {NoStop}%
\bibitem [{\citenamefont {Ono}\ \emph {et~al.}(2018)\citenamefont {Ono}, \citenamefont {Ishihara},\ and\ \citenamefont {Asada}}]{Ono:2018ybw}%
  \BibitemOpen
  \bibfield  {author} {\bibinfo {author} {\bibfnamefont {Toshiaki}\ \bibnamefont {Ono}}, \bibinfo {author} {\bibfnamefont {Asahi}\ \bibnamefont {Ishihara}}, \ and\ \bibinfo {author} {\bibfnamefont {Hideki}\ \bibnamefont {Asada}},\ }\bibfield  {title} {\enquote {\bibinfo {title} {{Deflection angle of light for an observer and source at finite distance from a rotating wormhole}},}\ }\href {\doibase 10.1103/PhysRevD.98.044047} {\bibfield  {journal} {\bibinfo  {journal} {Phys. Rev. D}\ }\textbf {\bibinfo {volume} {98}},\ \bibinfo {pages} {044047} (\bibinfo {year} {2018})},\ \Eprint {http://arxiv.org/abs/1806.05360} {arXiv:1806.05360 [gr-qc]} \BibitemShut {NoStop}%
\bibitem [{\citenamefont {Jusufi}\ \emph {et~al.}(2018)\citenamefont {Jusufi}, \citenamefont {\"Ovg\"un}, \citenamefont {Saavedra}, \citenamefont {V\'asquez},\ and\ \citenamefont {Gonz\'alez}}]{Jusufi:2018jof}%
  \BibitemOpen
  \bibfield  {author} {\bibinfo {author} {\bibfnamefont {Kimet}\ \bibnamefont {Jusufi}}, \bibinfo {author} {\bibfnamefont {Ali}\ \bibnamefont {\"Ovg\"un}}, \bibinfo {author} {\bibfnamefont {Joel}\ \bibnamefont {Saavedra}}, \bibinfo {author} {\bibfnamefont {Yerko}\ \bibnamefont {V\'asquez}}, \ and\ \bibinfo {author} {\bibfnamefont {P.~A.}\ \bibnamefont {Gonz\'alez}},\ }\bibfield  {title} {\enquote {\bibinfo {title} {{Deflection of light by rotating regular black holes using the Gauss-Bonnet theorem}},}\ }\href {\doibase 10.1103/PhysRevD.97.124024} {\bibfield  {journal} {\bibinfo  {journal} {Phys. Rev. D}\ }\textbf {\bibinfo {volume} {97}},\ \bibinfo {pages} {124024} (\bibinfo {year} {2018})},\ \Eprint {http://arxiv.org/abs/1804.00643} {arXiv:1804.00643 [gr-qc]} \BibitemShut {NoStop}%
\bibitem [{\citenamefont {\"Ovg\"un}(2018)}]{aliovgun}%
  \BibitemOpen
  \bibfield  {author} {\bibinfo {author} {\bibfnamefont {Ali}\ \bibnamefont {\"Ovg\"un}},\ }\bibfield  {title} {\enquote {\bibinfo {title} {{Light deflection by Damour-Solodukhin wormholes and Gauss-Bonnet theorem}},}\ }\href {\doibase 10.1103/PhysRevD.98.044033} {\bibfield  {journal} {\bibinfo  {journal} {Phys. Rev. D}\ }\textbf {\bibinfo {volume} {98}},\ \bibinfo {pages} {044033} (\bibinfo {year} {2018})},\ \Eprint {http://arxiv.org/abs/1805.06296} {arXiv:1805.06296 [gr-qc]} \BibitemShut {NoStop}%
\bibitem [{\citenamefont {\"Ovg\"un}(2019)}]{ali2}%
  \BibitemOpen
  \bibfield  {author} {\bibinfo {author} {\bibfnamefont {A.}~\bibnamefont {\"Ovg\"un}},\ }\bibfield  {title} {\enquote {\bibinfo {title} {{Weak field deflection angle by regular black holes with cosmic strings using the Gauss-Bonnet theorem}},}\ }\href {\doibase 10.1103/PhysRevD.99.104075} {\bibfield  {journal} {\bibinfo  {journal} {Phys. Rev. D}\ }\textbf {\bibinfo {volume} {99}},\ \bibinfo {pages} {104075} (\bibinfo {year} {2019})},\ \Eprint {http://arxiv.org/abs/1902.04411} {arXiv:1902.04411 [gr-qc]} \BibitemShut {NoStop}%
\bibitem [{\citenamefont {Javed}\ \emph {et~al.}(2019{\natexlab{a}})\citenamefont {Javed}, \citenamefont {Abbas},\ and\ \citenamefont {\"Ovg\"un}}]{javed}%
  \BibitemOpen
  \bibfield  {author} {\bibinfo {author} {\bibfnamefont {Wajiha}\ \bibnamefont {Javed}}, \bibinfo {author} {\bibfnamefont {Jameela}\ \bibnamefont {Abbas}}, \ and\ \bibinfo {author} {\bibfnamefont {Ali}\ \bibnamefont {\"Ovg\"un}},\ }\bibfield  {title} {\enquote {\bibinfo {title} {{Deflection angle of photon from magnetized black hole and effect of nonlinear electrodynamics}},}\ }\href {\doibase 10.1140/epjc/s10052-019-7208-3} {\bibfield  {journal} {\bibinfo  {journal} {Eur. Phys. J. C}\ }\textbf {\bibinfo {volume} {79}},\ \bibinfo {pages} {694} (\bibinfo {year} {2019}{\natexlab{a}})},\ \Eprint {http://arxiv.org/abs/1908.09632} {arXiv:1908.09632 [physics.gen-ph]} \BibitemShut {NoStop}%
\bibitem [{\citenamefont {Javed}\ \emph {et~al.}(2019{\natexlab{b}})\citenamefont {Javed}, \citenamefont {Babar},\ and\ \citenamefont {\"Ovg\"un}}]{babar}%
  \BibitemOpen
  \bibfield  {author} {\bibinfo {author} {\bibfnamefont {Wajiha}\ \bibnamefont {Javed}}, \bibinfo {author} {\bibfnamefont {Rimsha}\ \bibnamefont {Babar}}, \ and\ \bibinfo {author} {\bibfnamefont {Al\"\i{}}\ \bibnamefont {\"Ovg\"un}},\ }\bibfield  {title} {\enquote {\bibinfo {title} {{Effect of the dilaton field and plasma medium on deflection angle by black holes in Einstein-Maxwell-dilaton-axion theory}},}\ }\href {\doibase 10.1103/PhysRevD.100.104032} {\bibfield  {journal} {\bibinfo  {journal} {Phys. Rev. D}\ }\textbf {\bibinfo {volume} {100}},\ \bibinfo {pages} {104032} (\bibinfo {year} {2019}{\natexlab{b}})},\ \Eprint {http://arxiv.org/abs/1910.11697} {arXiv:1910.11697 [gr-qc]} \BibitemShut {NoStop}%
\bibitem [{\citenamefont {de~Leon}\ and\ \citenamefont {Vega}(2019)}]{deLeon:2019qnp}%
  \BibitemOpen
  \bibfield  {author} {\bibinfo {author} {\bibfnamefont {Karlo}\ \bibnamefont {de~Leon}}\ and\ \bibinfo {author} {\bibfnamefont {Ian}\ \bibnamefont {Vega}},\ }\bibfield  {title} {\enquote {\bibinfo {title} {{Weak gravitational deflection by two-power-law densities using the Gauss-Bonnet theorem}},}\ }\href {\doibase 10.1103/PhysRevD.99.124007} {\bibfield  {journal} {\bibinfo  {journal} {Phys. Rev. D}\ }\textbf {\bibinfo {volume} {99}},\ \bibinfo {pages} {124007} (\bibinfo {year} {2019})},\ \Eprint {http://arxiv.org/abs/1903.06951} {arXiv:1903.06951 [gr-qc]} \BibitemShut {NoStop}%
\bibitem [{\citenamefont {Kumar}\ \emph {et~al.}(2019)\citenamefont {Kumar}, \citenamefont {Ghosh},\ and\ \citenamefont {Wang}}]{Kumar:2019pjp}%
  \BibitemOpen
  \bibfield  {author} {\bibinfo {author} {\bibfnamefont {Rahul}\ \bibnamefont {Kumar}}, \bibinfo {author} {\bibfnamefont {Sushant~G.}\ \bibnamefont {Ghosh}}, \ and\ \bibinfo {author} {\bibfnamefont {Anzhong}\ \bibnamefont {Wang}},\ }\bibfield  {title} {\enquote {\bibinfo {title} {{Shadow cast and deflection of light by charged rotating regular black holes}},}\ }\href {\doibase 10.1103/PhysRevD.100.124024} {\bibfield  {journal} {\bibinfo  {journal} {Phys. Rev. D}\ }\textbf {\bibinfo {volume} {100}},\ \bibinfo {pages} {124024} (\bibinfo {year} {2019})},\ \Eprint {http://arxiv.org/abs/1912.05154} {arXiv:1912.05154 [gr-qc]} \BibitemShut {NoStop}%
\bibitem [{\citenamefont {Javed}\ \emph {et~al.}(2019{\natexlab{c}})\citenamefont {Javed}, \citenamefont {Babar},\ and\ \citenamefont {\"Ovg\"un}}]{javed1}%
  \BibitemOpen
  \bibfield  {author} {\bibinfo {author} {\bibfnamefont {Wajiha}\ \bibnamefont {Javed}}, \bibinfo {author} {\bibfnamefont {Rimsha}\ \bibnamefont {Babar}}, \ and\ \bibinfo {author} {\bibfnamefont {Ali}\ \bibnamefont {\"Ovg\"un}},\ }\bibfield  {title} {\enquote {\bibinfo {title} {{The effect of the Brane-Dicke coupling parameter on weak gravitational lensing by wormholes and naked singularities}},}\ }\href {\doibase 10.1103/PhysRevD.99.084012} {\bibfield  {journal} {\bibinfo  {journal} {Phys. Rev. D}\ }\textbf {\bibinfo {volume} {99}},\ \bibinfo {pages} {084012} (\bibinfo {year} {2019}{\natexlab{c}})},\ \Eprint {http://arxiv.org/abs/1903.11657} {arXiv:1903.11657 [gr-qc]} \BibitemShut {NoStop}%
\bibitem [{\citenamefont {Javed}\ \emph {et~al.}(2022)\citenamefont {Javed}, \citenamefont {Aqib},\ and\ \citenamefont {\"Ovg\"un}}]{aqib}%
  \BibitemOpen
  \bibfield  {author} {\bibinfo {author} {\bibfnamefont {Wajiha}\ \bibnamefont {Javed}}, \bibinfo {author} {\bibfnamefont {Muhammad}\ \bibnamefont {Aqib}}, \ and\ \bibinfo {author} {\bibfnamefont {Ali}\ \bibnamefont {\"Ovg\"un}},\ }\bibfield  {title} {\enquote {\bibinfo {title} {{Effect of the magnetic charge on weak deflection angle and greybody bound of the black hole in Einstein-Gauss-Bonnet gravity}},}\ }\href {\doibase 10.1016/j.physletb.2022.137114} {\bibfield  {journal} {\bibinfo  {journal} {Phys. Lett. B}\ }\textbf {\bibinfo {volume} {829}},\ \bibinfo {pages} {137114} (\bibinfo {year} {2022})},\ \Eprint {http://arxiv.org/abs/2204.07864} {arXiv:2204.07864 [gr-qc]} \BibitemShut {NoStop}%
\bibitem [{\citenamefont {Mustafa}\ \emph {et~al.}(2024)\citenamefont {Mustafa}, \citenamefont {Ditta}, \citenamefont {Javed}, \citenamefont {Atamurotov}, \citenamefont {Hussain},\ and\ \citenamefont {Ahmedov}}]{mustafa}%
  \BibitemOpen
  \bibfield  {author} {\bibinfo {author} {\bibfnamefont {G.}~\bibnamefont {Mustafa}}, \bibinfo {author} {\bibfnamefont {Allah}\ \bibnamefont {Ditta}}, \bibinfo {author} {\bibfnamefont {Faisal}\ \bibnamefont {Javed}}, \bibinfo {author} {\bibfnamefont {Farruh}\ \bibnamefont {Atamurotov}}, \bibinfo {author} {\bibfnamefont {Ibrar}\ \bibnamefont {Hussain}}, \ and\ \bibinfo {author} {\bibfnamefont {Bobomurat}\ \bibnamefont {Ahmedov}},\ }\bibfield  {title} {\enquote {\bibinfo {title} {{Probing a black hole in Starobinsky-Bel-Robinson gravity with thermodynamical analysis, effective force and gravitational weak lensing}},}\ }\href {\doibase 10.1016/j.cjph.2024.04.038} {\bibfield  {journal} {\bibinfo  {journal} {Chin. J. Phys.}\ }\textbf {\bibinfo {volume} {90}},\ \bibinfo {pages} {494--508} (\bibinfo {year} {2024})},\ \Eprint {http://arxiv.org/abs/2401.08254} {arXiv:2401.08254 [gr-qc]} \BibitemShut {NoStop}%
\bibitem [{\citenamefont {Gao}\ and\ \citenamefont {Liu}(2024)}]{Gao:2023sla}%
  \BibitemOpen
  \bibfield  {author} {\bibinfo {author} {\bibfnamefont {Ke}~\bibnamefont {Gao}}\ and\ \bibinfo {author} {\bibfnamefont {Lei-Hua}\ \bibnamefont {Liu}},\ }\bibfield  {title} {\enquote {\bibinfo {title} {{Microlensing and event rate of static spherically symmetric wormhole}},}\ }\href {\doibase 10.1016/j.physletb.2024.139019} {\bibfield  {journal} {\bibinfo  {journal} {Phys. Lett. B}\ }\textbf {\bibinfo {volume} {858}},\ \bibinfo {pages} {139019} (\bibinfo {year} {2024})},\ \Eprint {http://arxiv.org/abs/2303.11134} {arXiv:2303.11134 [gr-qc]} \BibitemShut {NoStop}%
\bibitem [{\citenamefont {Gao}\ \emph {et~al.}(2023)\citenamefont {Gao}, \citenamefont {Liu},\ and\ \citenamefont {Zhu}}]{Gao:2022cds}%
  \BibitemOpen
  \bibfield  {author} {\bibinfo {author} {\bibfnamefont {Ke}~\bibnamefont {Gao}}, \bibinfo {author} {\bibfnamefont {Lei-Hua}\ \bibnamefont {Liu}}, \ and\ \bibinfo {author} {\bibfnamefont {Mian}\ \bibnamefont {Zhu}},\ }\bibfield  {title} {\enquote {\bibinfo {title} {{Microlensing effects of wormholes associated to blackhole spacetimes}},}\ }\href {\doibase 10.1016/j.dark.2023.101254} {\bibfield  {journal} {\bibinfo  {journal} {Phys. Dark Univ.}\ }\textbf {\bibinfo {volume} {41}},\ \bibinfo {pages} {101254} (\bibinfo {year} {2023})},\ \Eprint {http://arxiv.org/abs/2211.17065} {arXiv:2211.17065 [gr-qc]} \BibitemShut {NoStop}%
\bibitem [{\citenamefont {Qiao}\ and\ \citenamefont {Zhou}(2023)}]{Qiao:2022nic}%
  \BibitemOpen
  \bibfield  {author} {\bibinfo {author} {\bibfnamefont {Chen-Kai}\ \bibnamefont {Qiao}}\ and\ \bibinfo {author} {\bibfnamefont {Mi}~\bibnamefont {Zhou}},\ }\bibfield  {title} {\enquote {\bibinfo {title} {{Gravitational lensing of Schwarzschild and charged black holes immersed in perfect fluid dark matter halo}},}\ }\href {\doibase 10.1088/1475-7516/2023/12/005} {\bibfield  {journal} {\bibinfo  {journal} {JCAP}\ }\textbf {\bibinfo {volume} {12}},\ \bibinfo {pages} {005} (\bibinfo {year} {2023})},\ \Eprint {http://arxiv.org/abs/2212.13311} {arXiv:2212.13311 [gr-qc]} \BibitemShut {NoStop}%
\bibitem [{\citenamefont {Huang}\ \emph {et~al.}(2023)\citenamefont {Huang}, \citenamefont {Sun},\ and\ \citenamefont {Cao}}]{Huang:2022gon}%
  \BibitemOpen
  \bibfield  {author} {\bibinfo {author} {\bibfnamefont {Yang}\ \bibnamefont {Huang}}, \bibinfo {author} {\bibfnamefont {Bing}\ \bibnamefont {Sun}}, \ and\ \bibinfo {author} {\bibfnamefont {Zhoujian}\ \bibnamefont {Cao}},\ }\bibfield  {title} {\enquote {\bibinfo {title} {{Extending Gibbons-Werner method to bound orbits of massive particles}},}\ }\href {\doibase 10.1103/PhysRevD.107.104046} {\bibfield  {journal} {\bibinfo  {journal} {Phys. Rev. D}\ }\textbf {\bibinfo {volume} {107}},\ \bibinfo {pages} {104046} (\bibinfo {year} {2023})},\ \Eprint {http://arxiv.org/abs/2212.04251} {arXiv:2212.04251 [gr-qc]} \BibitemShut {NoStop}%
\bibitem [{\citenamefont {Javed}\ \emph {et~al.}(2023{\natexlab{a}})\citenamefont {Javed}, \citenamefont {Atique}, \citenamefont {Pantig},\ and\ \citenamefont {\"Ovg\"un}}]{Javed:2023iih}%
  \BibitemOpen
  \bibfield  {author} {\bibinfo {author} {\bibfnamefont {Wajiha}\ \bibnamefont {Javed}}, \bibinfo {author} {\bibfnamefont {Mehak}\ \bibnamefont {Atique}}, \bibinfo {author} {\bibfnamefont {Reggie~C.}\ \bibnamefont {Pantig}}, \ and\ \bibinfo {author} {\bibfnamefont {Ali}\ \bibnamefont {\"Ovg\"un}},\ }\bibfield  {title} {\enquote {\bibinfo {title} {{Weak Deflection Angle, Hawking Radiation and Greybody Bound of Reissner\textendash{}Nordstr\"om Black Hole Corrected by Bounce Parameter}},}\ }\href {\doibase 10.3390/sym15010148} {\bibfield  {journal} {\bibinfo  {journal} {Symmetry}\ }\textbf {\bibinfo {volume} {15}},\ \bibinfo {pages} {148} (\bibinfo {year} {2023}{\natexlab{a}})},\ \Eprint {http://arxiv.org/abs/2301.01855} {arXiv:2301.01855 [gr-qc]} \BibitemShut {NoStop}%
\bibitem [{\citenamefont {Javed}\ \emph {et~al.}(2023{\natexlab{b}})\citenamefont {Javed}, \citenamefont {Atique}, \citenamefont {Pantig},\ and\ \citenamefont {\"Ovg\"un}}]{Javed:2022psa}%
  \BibitemOpen
  \bibfield  {author} {\bibinfo {author} {\bibfnamefont {Wajiha}\ \bibnamefont {Javed}}, \bibinfo {author} {\bibfnamefont {Mehak}\ \bibnamefont {Atique}}, \bibinfo {author} {\bibfnamefont {Reggie~C.}\ \bibnamefont {Pantig}}, \ and\ \bibinfo {author} {\bibfnamefont {Ali}\ \bibnamefont {\"Ovg\"un}},\ }\bibfield  {title} {\enquote {\bibinfo {title} {{Weak lensing, Hawking radiation and greybody factor bound by a charged black holes with non-linear electrodynamics corrections}},}\ }\href {\doibase 10.1142/S0219887823500408} {\bibfield  {journal} {\bibinfo  {journal} {Int. J. Geom. Meth. Mod. Phys.}\ }\textbf {\bibinfo {volume} {20}},\ \bibinfo {pages} {2350040} (\bibinfo {year} {2023}{\natexlab{b}})}\BibitemShut {NoStop}%
\bibitem [{\citenamefont {Okyay}\ and\ \citenamefont {\"Ovg\"un}(2022)}]{Okyay:2021nnh}%
  \BibitemOpen
  \bibfield  {author} {\bibinfo {author} {\bibfnamefont {Mert}\ \bibnamefont {Okyay}}\ and\ \bibinfo {author} {\bibfnamefont {Ali}\ \bibnamefont {\"Ovg\"un}},\ }\bibfield  {title} {\enquote {\bibinfo {title} {{Nonlinear electrodynamics effects on the black hole shadow, deflection angle, quasinormal modes and greybody factors}},}\ }\href {\doibase 10.1088/1475-7516/2022/01/009} {\bibfield  {journal} {\bibinfo  {journal} {JCAP}\ }\textbf {\bibinfo {volume} {01}},\ \bibinfo {pages} {009} (\bibinfo {year} {2022})},\ \Eprint {http://arxiv.org/abs/2108.07766} {arXiv:2108.07766 [gr-qc]} \BibitemShut {NoStop}%
\bibitem [{\citenamefont {Rayimbaev}\ \emph {et~al.}(2023)\citenamefont {Rayimbaev}, \citenamefont {Pantig}, \citenamefont {\"Ovg\"un}, \citenamefont {Abdujabbarov},\ and\ \citenamefont {Demir}}]{Rayimbaev:2022hca}%
  \BibitemOpen
  \bibfield  {author} {\bibinfo {author} {\bibfnamefont {Javlon}\ \bibnamefont {Rayimbaev}}, \bibinfo {author} {\bibfnamefont {Reggie~C.}\ \bibnamefont {Pantig}}, \bibinfo {author} {\bibfnamefont {Ali}\ \bibnamefont {\"Ovg\"un}}, \bibinfo {author} {\bibfnamefont {Ahmadjon}\ \bibnamefont {Abdujabbarov}}, \ and\ \bibinfo {author} {\bibfnamefont {Durmu\c{s}}\ \bibnamefont {Demir}},\ }\bibfield  {title} {\enquote {\bibinfo {title} {{Quasiperiodic oscillations, weak field lensing and shadow cast around black holes in Symmergent gravity}},}\ }\href {\doibase 10.1016/j.aop.2023.169335} {\bibfield  {journal} {\bibinfo  {journal} {Annals Phys.}\ }\textbf {\bibinfo {volume} {454}},\ \bibinfo {pages} {169335} (\bibinfo {year} {2023})},\ \Eprint {http://arxiv.org/abs/2206.06599} {arXiv:2206.06599 [gr-qc]} \BibitemShut {NoStop}%
\bibitem [{\citenamefont {Rindler}\ and\ \citenamefont {Ishak}(2007)}]{Rindler:2007zz}%
  \BibitemOpen
  \bibfield  {author} {\bibinfo {author} {\bibfnamefont {Wolfgang}\ \bibnamefont {Rindler}}\ and\ \bibinfo {author} {\bibfnamefont {Mustapha}\ \bibnamefont {Ishak}},\ }\bibfield  {title} {\enquote {\bibinfo {title} {{Contribution of the cosmological constant to the relativistic bending of light revisited}},}\ }\href {\doibase 10.1103/PhysRevD.76.043006} {\bibfield  {journal} {\bibinfo  {journal} {Phys. Rev. D}\ }\textbf {\bibinfo {volume} {76}},\ \bibinfo {pages} {043006} (\bibinfo {year} {2007})},\ \Eprint {http://arxiv.org/abs/0709.2948} {arXiv:0709.2948 [astro-ph]} \BibitemShut {NoStop}%
\bibitem [{\citenamefont {Park}(2008)}]{Park:2008ih}%
  \BibitemOpen
  \bibfield  {author} {\bibinfo {author} {\bibfnamefont {Minjoon}\ \bibnamefont {Park}},\ }\bibfield  {title} {\enquote {\bibinfo {title} {{Rigorous Approach to the Gravitational Lensing}},}\ }\href {\doibase 10.1103/PhysRevD.78.023014} {\bibfield  {journal} {\bibinfo  {journal} {Phys. Rev. D}\ }\textbf {\bibinfo {volume} {78}},\ \bibinfo {pages} {023014} (\bibinfo {year} {2008})},\ \Eprint {http://arxiv.org/abs/0804.4331} {arXiv:0804.4331 [astro-ph]} \BibitemShut {NoStop}%
\bibitem [{\citenamefont {Sereno}(2009)}]{Sereno:2008kk}%
  \BibitemOpen
  \bibfield  {author} {\bibinfo {author} {\bibfnamefont {M.}~\bibnamefont {Sereno}},\ }\bibfield  {title} {\enquote {\bibinfo {title} {{The role of Lambda in the cosmological lens equation}},}\ }\href {\doibase 10.1103/PhysRevLett.102.021301} {\bibfield  {journal} {\bibinfo  {journal} {Phys. Rev. Lett.}\ }\textbf {\bibinfo {volume} {102}},\ \bibinfo {pages} {021301} (\bibinfo {year} {2009})},\ \Eprint {http://arxiv.org/abs/0807.5123} {arXiv:0807.5123 [astro-ph]} \BibitemShut {NoStop}%
\bibitem [{\citenamefont {Simpson}\ \emph {et~al.}(2010)\citenamefont {Simpson}, \citenamefont {Peacock},\ and\ \citenamefont {Heavens}}]{peacock}%
  \BibitemOpen
  \bibfield  {author} {\bibinfo {author} {\bibfnamefont {Fergus}\ \bibnamefont {Simpson}}, \bibinfo {author} {\bibfnamefont {John~A.}\ \bibnamefont {Peacock}}, \ and\ \bibinfo {author} {\bibfnamefont {Alan~F.}\ \bibnamefont {Heavens}},\ }\bibfield  {title} {\enquote {\bibinfo {title} {{On lensing by a cosmological constant}},}\ }\href {\doibase 10.1111/j.1365-2966.2009.16032.x} {\bibfield  {journal} {\bibinfo  {journal} {Mon. Not. Roy. Astron. Soc.}\ }\textbf {\bibinfo {volume} {402}},\ \bibinfo {pages} {2009} (\bibinfo {year} {2010})},\ \Eprint {http://arxiv.org/abs/0809.1819} {arXiv:0809.1819 [astro-ph]} \BibitemShut {NoStop}%
\bibitem [{\citenamefont {Li}\ and\ \citenamefont {Zhou}(2021)}]{Li:2020zxi}%
  \BibitemOpen
  \bibfield  {author} {\bibinfo {author} {\bibfnamefont {Zonghai}\ \bibnamefont {Li}}\ and\ \bibinfo {author} {\bibfnamefont {Tao}\ \bibnamefont {Zhou}},\ }\bibfield  {title} {\enquote {\bibinfo {title} {{Kerr black hole surrounded by a cloud of strings and its weak gravitational lensing in Rastall gravity}},}\ }\href {\doibase 10.1103/PhysRevD.104.104044} {\bibfield  {journal} {\bibinfo  {journal} {Phys. Rev. D}\ }\textbf {\bibinfo {volume} {104}},\ \bibinfo {pages} {104044} (\bibinfo {year} {2021})},\ \Eprint {http://arxiv.org/abs/2001.01642} {arXiv:2001.01642 [gr-qc]} \BibitemShut {NoStop}%
\bibitem [{\citenamefont {Aad}\ \emph {et~al.}(2024)\citenamefont {Aad} \emph {et~al.}}]{ATLAS:2024mih}%
  \BibitemOpen
  \bibfield  {author} {\bibinfo {author} {\bibfnamefont {Georges}\ \bibnamefont {Aad}} \emph {et~al.} (\bibinfo {collaboration} {ATLAS}),\ }\bibfield  {title} {\enquote {\bibinfo {title} {{Search for flavour-changing neutral-current couplings between the top quark and the Higgs boson in multi-lepton final states in 13~TeV pp collisions with the ATLAS detector}},}\ }\href {\doibase 10.1140/epjc/s10052-024-12994-1} {\bibfield  {journal} {\bibinfo  {journal} {Eur. Phys. J. C}\ }\textbf {\bibinfo {volume} {84}},\ \bibinfo {pages} {757} (\bibinfo {year} {2024})},\ \Eprint {http://arxiv.org/abs/2404.02123} {arXiv:2404.02123 [hep-ex]} \BibitemShut {NoStop}%
\bibitem [{\citenamefont {Li}\ and\ \citenamefont {\"Ovg\"un}(2020)}]{Li:2020dln}%
  \BibitemOpen
  \bibfield  {author} {\bibinfo {author} {\bibfnamefont {Zonghai}\ \bibnamefont {Li}}\ and\ \bibinfo {author} {\bibfnamefont {Ali}\ \bibnamefont {\"Ovg\"un}},\ }\bibfield  {title} {\enquote {\bibinfo {title} {{Finite-distance gravitational deflection of massive particles by a Kerr-like black hole in the bumblebee gravity model}},}\ }\href {\doibase 10.1103/PhysRevD.101.024040} {\bibfield  {journal} {\bibinfo  {journal} {Phys. Rev. D}\ }\textbf {\bibinfo {volume} {101}},\ \bibinfo {pages} {024040} (\bibinfo {year} {2020})},\ \Eprint {http://arxiv.org/abs/2001.02074} {arXiv:2001.02074 [gr-qc]} \BibitemShut {NoStop}%
\bibitem [{\citenamefont {Arakida}(2018{\natexlab{b}})}]{Arakida:2017hrm}%
  \BibitemOpen
  \bibfield  {author} {\bibinfo {author} {\bibfnamefont {Hideyoshi}\ \bibnamefont {Arakida}},\ }\bibfield  {title} {\enquote {\bibinfo {title} {{Light deflection and Gauss\textendash{}Bonnet theorem: definition of total deflection angle and its applications}},}\ }\href {\doibase 10.1007/s10714-018-2368-2} {\bibfield  {journal} {\bibinfo  {journal} {Gen. Rel. Grav.}\ }\textbf {\bibinfo {volume} {50}},\ \bibinfo {pages} {48} (\bibinfo {year} {2018}{\natexlab{b}})},\ \Eprint {http://arxiv.org/abs/1708.04011} {arXiv:1708.04011 [gr-qc]} \BibitemShut {NoStop}%
\bibitem [{\citenamefont {Takizawa}\ \emph {et~al.}(2020)\citenamefont {Takizawa}, \citenamefont {Ono},\ and\ \citenamefont {Asada}}]{Takizawa:2020egm}%
  \BibitemOpen
  \bibfield  {author} {\bibinfo {author} {\bibfnamefont {Keita}\ \bibnamefont {Takizawa}}, \bibinfo {author} {\bibfnamefont {Toshiaki}\ \bibnamefont {Ono}}, \ and\ \bibinfo {author} {\bibfnamefont {Hideki}\ \bibnamefont {Asada}},\ }\bibfield  {title} {\enquote {\bibinfo {title} {{Gravitational deflection angle of light: Definition by an observer and its application to an asymptotically nonflat spacetime}},}\ }\href {\doibase 10.1103/PhysRevD.101.104032} {\bibfield  {journal} {\bibinfo  {journal} {Phys. Rev. D}\ }\textbf {\bibinfo {volume} {101}},\ \bibinfo {pages} {104032} (\bibinfo {year} {2020})},\ \Eprint {http://arxiv.org/abs/2001.03290} {arXiv:2001.03290 [gr-qc]} \BibitemShut {NoStop}%
\bibitem [{\citenamefont {Luminet}(1979)}]{Luminet:1979nyg}%
  \BibitemOpen
  \bibfield  {author} {\bibinfo {author} {\bibfnamefont {J.~P.}\ \bibnamefont {Luminet}},\ }\bibfield  {title} {\enquote {\bibinfo {title} {{Image of a spherical black hole with thin accretion disk}},}\ }\href@noop {} {\bibfield  {journal} {\bibinfo  {journal} {Astron. Astrophys.}\ }\textbf {\bibinfo {volume} {75}},\ \bibinfo {pages} {228--235} (\bibinfo {year} {1979})}\BibitemShut {NoStop}%
\bibitem [{\citenamefont {Akiyama}\ \emph {et~al.}(2019{\natexlab{a}})\citenamefont {Akiyama} \emph {et~al.}}]{EventHorizonTelescope:2019dse}%
  \BibitemOpen
  \bibfield  {author} {\bibinfo {author} {\bibfnamefont {Kazunori}\ \bibnamefont {Akiyama}} \emph {et~al.} (\bibinfo {collaboration} {Event Horizon Telescope}),\ }\bibfield  {title} {\enquote {\bibinfo {title} {{First M87 Event Horizon Telescope Results. I. The Shadow of the Supermassive Black Hole}},}\ }\href {\doibase 10.3847/2041-8213/ab0ec7} {\bibfield  {journal} {\bibinfo  {journal} {Astrophys. J. Lett.}\ }\textbf {\bibinfo {volume} {875}},\ \bibinfo {pages} {L1} (\bibinfo {year} {2019}{\natexlab{a}})},\ \Eprint {http://arxiv.org/abs/1906.11238} {arXiv:1906.11238 [astro-ph.GA]} \BibitemShut {NoStop}%
\bibitem [{\citenamefont {Akiyama}\ \emph {et~al.}(2019{\natexlab{b}})\citenamefont {Akiyama} \emph {et~al.}}]{EventHorizonTelescope:2019ths}%
  \BibitemOpen
  \bibfield  {author} {\bibinfo {author} {\bibfnamefont {Kazunori}\ \bibnamefont {Akiyama}} \emph {et~al.} (\bibinfo {collaboration} {Event Horizon Telescope}),\ }\bibfield  {title} {\enquote {\bibinfo {title} {{First M87 Event Horizon Telescope Results. IV. Imaging the Central Supermassive Black Hole}},}\ }\href {\doibase 10.3847/2041-8213/ab0e85} {\bibfield  {journal} {\bibinfo  {journal} {Astrophys. J. Lett.}\ }\textbf {\bibinfo {volume} {875}},\ \bibinfo {pages} {L4} (\bibinfo {year} {2019}{\natexlab{b}})},\ \Eprint {http://arxiv.org/abs/1906.11241} {arXiv:1906.11241 [astro-ph.GA]} \BibitemShut {NoStop}%
\bibitem [{\citenamefont {Akiyama}\ \emph {et~al.}(2019{\natexlab{c}})\citenamefont {Akiyama} \emph {et~al.}}]{EventHorizonTelescope:2019ggy}%
  \BibitemOpen
  \bibfield  {author} {\bibinfo {author} {\bibfnamefont {Kazunori}\ \bibnamefont {Akiyama}} \emph {et~al.} (\bibinfo {collaboration} {Event Horizon Telescope}),\ }\bibfield  {title} {\enquote {\bibinfo {title} {{First M87 Event Horizon Telescope Results. VI. The Shadow and Mass of the Central Black Hole}},}\ }\href {\doibase 10.3847/2041-8213/ab1141} {\bibfield  {journal} {\bibinfo  {journal} {Astrophys. J. Lett.}\ }\textbf {\bibinfo {volume} {875}},\ \bibinfo {pages} {L6} (\bibinfo {year} {2019}{\natexlab{c}})},\ \Eprint {http://arxiv.org/abs/1906.11243} {arXiv:1906.11243 [astro-ph.GA]} \BibitemShut {NoStop}%
\bibitem [{\citenamefont {Akiyama}\ \emph {et~al.}(2022{\natexlab{a}})\citenamefont {Akiyama} \emph {et~al.}}]{EventHorizonTelescope:2022wkp}%
  \BibitemOpen
  \bibfield  {author} {\bibinfo {author} {\bibfnamefont {Kazunori}\ \bibnamefont {Akiyama}} \emph {et~al.} (\bibinfo {collaboration} {Event Horizon Telescope}),\ }\bibfield  {title} {\enquote {\bibinfo {title} {{First Sagittarius A* Event Horizon Telescope Results. I. The Shadow of the Supermassive Black Hole in the Center of the Milky Way}},}\ }\href {\doibase 10.3847/2041-8213/ac6674} {\bibfield  {journal} {\bibinfo  {journal} {Astrophys. J. Lett.}\ }\textbf {\bibinfo {volume} {930}},\ \bibinfo {pages} {L12} (\bibinfo {year} {2022}{\natexlab{a}})},\ \Eprint {http://arxiv.org/abs/2311.08680} {arXiv:2311.08680 [astro-ph.HE]} \BibitemShut {NoStop}%
\bibitem [{\citenamefont {Akiyama}\ \emph {et~al.}(2022{\natexlab{b}})\citenamefont {Akiyama} \emph {et~al.}}]{EventHorizonTelescope:2022wok}%
  \BibitemOpen
  \bibfield  {author} {\bibinfo {author} {\bibfnamefont {Kazunori}\ \bibnamefont {Akiyama}} \emph {et~al.} (\bibinfo {collaboration} {Event Horizon Telescope}),\ }\bibfield  {title} {\enquote {\bibinfo {title} {{First Sagittarius A* Event Horizon Telescope Results. III. Imaging of the Galactic Center Supermassive Black Hole}},}\ }\href {\doibase 10.3847/2041-8213/ac6429} {\bibfield  {journal} {\bibinfo  {journal} {Astrophys. J. Lett.}\ }\textbf {\bibinfo {volume} {930}},\ \bibinfo {pages} {L14} (\bibinfo {year} {2022}{\natexlab{b}})},\ \Eprint {http://arxiv.org/abs/2311.09479} {arXiv:2311.09479 [astro-ph.HE]} \BibitemShut {NoStop}%
\bibitem [{\citenamefont {Akiyama}\ \emph {et~al.}(2022{\natexlab{c}})\citenamefont {Akiyama} \emph {et~al.}}]{EventHorizonTelescope:2022xqj}%
  \BibitemOpen
  \bibfield  {author} {\bibinfo {author} {\bibfnamefont {Kazunori}\ \bibnamefont {Akiyama}} \emph {et~al.} (\bibinfo {collaboration} {Event Horizon Telescope}),\ }\bibfield  {title} {\enquote {\bibinfo {title} {{First Sagittarius A* Event Horizon Telescope Results. VI. Testing the Black Hole Metric}},}\ }\href {\doibase 10.3847/2041-8213/ac6756} {\bibfield  {journal} {\bibinfo  {journal} {Astrophys. J. Lett.}\ }\textbf {\bibinfo {volume} {930}},\ \bibinfo {pages} {L17} (\bibinfo {year} {2022}{\natexlab{c}})},\ \Eprint {http://arxiv.org/abs/2311.09484} {arXiv:2311.09484 [astro-ph.HE]} \BibitemShut {NoStop}%
\bibitem [{\citenamefont {Yan}(2024)}]{Yan:2024rsx}%
  \BibitemOpen
  \bibfield  {author} {\bibinfo {author} {\bibfnamefont {Zening}\ \bibnamefont {Yan}},\ }\href@noop {} {\enquote {\bibinfo {title} {{Employing shadow radius to constrain extra dimensions in black string space-time with dark matter halo}},}\ } (\bibinfo {year} {2024}),\ \Eprint {http://arxiv.org/abs/2412.20109} {arXiv:2412.20109 [gr-qc]} \BibitemShut {NoStop}%
\bibitem [{\citenamefont {Yang}(2024)}]{Yang:2024ulu}%
  \BibitemOpen
  \bibfield  {author} {\bibinfo {author} {\bibfnamefont {Xuetao}\ \bibnamefont {Yang}},\ }\bibfield  {title} {\enquote {\bibinfo {title} {{Observational appearance of the spherically symmetric black hole in PFDM}},}\ }\href {\doibase 10.1016/j.dark.2024.101467} {\bibfield  {journal} {\bibinfo  {journal} {Phys. Dark Univ.}\ }\textbf {\bibinfo {volume} {44}},\ \bibinfo {pages} {101467} (\bibinfo {year} {2024})}\BibitemShut {NoStop}%
\bibitem [{\citenamefont {Yang}\ \emph {et~al.}(2024)\citenamefont {Yang}, \citenamefont {Liu}, \citenamefont {\"Ovg\"un}, \citenamefont {Lambiase},\ and\ \citenamefont {Long}}]{Yang:2023tip}%
  \BibitemOpen
  \bibfield  {author} {\bibinfo {author} {\bibfnamefont {Yi}~\bibnamefont {Yang}}, \bibinfo {author} {\bibfnamefont {Dong}\ \bibnamefont {Liu}}, \bibinfo {author} {\bibfnamefont {Ali}\ \bibnamefont {\"Ovg\"un}}, \bibinfo {author} {\bibfnamefont {Gaetano}\ \bibnamefont {Lambiase}}, \ and\ \bibinfo {author} {\bibfnamefont {Zheng-Wen}\ \bibnamefont {Long}},\ }\bibfield  {title} {\enquote {\bibinfo {title} {{Black hole surrounded by the pseudo-isothermal dark matter halo}},}\ }\href {\doibase 10.1140/epjc/s10052-024-12412-6} {\bibfield  {journal} {\bibinfo  {journal} {Eur. Phys. J. C}\ }\textbf {\bibinfo {volume} {84}},\ \bibinfo {pages} {63} (\bibinfo {year} {2024})},\ \Eprint {http://arxiv.org/abs/2308.05544} {arXiv:2308.05544 [gr-qc]} \BibitemShut {NoStop}%
\bibitem [{\citenamefont {Chakhchi}\ \emph {et~al.}(2024)\citenamefont {Chakhchi}, \citenamefont {El~Moumni},\ and\ \citenamefont {Masmar}}]{Chakhchi:2024tzo}%
  \BibitemOpen
  \bibfield  {author} {\bibinfo {author} {\bibfnamefont {L.}~\bibnamefont {Chakhchi}}, \bibinfo {author} {\bibfnamefont {H.}~\bibnamefont {El~Moumni}}, \ and\ \bibinfo {author} {\bibfnamefont {K.}~\bibnamefont {Masmar}},\ }\bibfield  {title} {\enquote {\bibinfo {title} {{Signatures of the accelerating black holes with a cosmological constant from the Sgr A\ensuremath{\star} and M87\ensuremath{\star} shadow prospects}},}\ }\href {\doibase 10.1016/j.dark.2024.101501} {\bibfield  {journal} {\bibinfo  {journal} {Phys. Dark Univ.}\ }\textbf {\bibinfo {volume} {44}},\ \bibinfo {pages} {101501} (\bibinfo {year} {2024})},\ \Eprint {http://arxiv.org/abs/2403.09756} {arXiv:2403.09756 [gr-qc]} \BibitemShut {NoStop}%
\bibitem [{\citenamefont {Jafarzade}\ \emph {et~al.}(2024)\citenamefont {Jafarzade}, \citenamefont {Eslam~Panah},\ and\ \citenamefont {Rodrigues}}]{Jafarzade:2024knc}%
  \BibitemOpen
  \bibfield  {author} {\bibinfo {author} {\bibfnamefont {Kh.}\ \bibnamefont {Jafarzade}}, \bibinfo {author} {\bibfnamefont {B.}~\bibnamefont {Eslam~Panah}}, \ and\ \bibinfo {author} {\bibfnamefont {M.~E.}\ \bibnamefont {Rodrigues}},\ }\bibfield  {title} {\enquote {\bibinfo {title} {{Thermodynamics and optical properties of phantom AdS black holes in massive gravity}},}\ }\href {\doibase 10.1088/1361-6382/ad242e} {\bibfield  {journal} {\bibinfo  {journal} {Class. Quant. Grav.}\ }\textbf {\bibinfo {volume} {41}},\ \bibinfo {pages} {065007} (\bibinfo {year} {2024})},\ \Eprint {http://arxiv.org/abs/2402.08704} {arXiv:2402.08704 [gr-qc]} \BibitemShut {NoStop}%
\bibitem [{\citenamefont {\"Ovg\"un}\ \emph {et~al.}(2023)\citenamefont {\"Ovg\"un}, \citenamefont {Sese},\ and\ \citenamefont {Pantig}}]{Ovgun:2023wmc}%
  \BibitemOpen
  \bibfield  {author} {\bibinfo {author} {\bibfnamefont {Ali}\ \bibnamefont {\"Ovg\"un}}, \bibinfo {author} {\bibfnamefont {Lemuel John~F.}\ \bibnamefont {Sese}}, \ and\ \bibinfo {author} {\bibfnamefont {Reggie~C.}\ \bibnamefont {Pantig}},\ }\bibfield  {title} {\enquote {\bibinfo {title} {{Constraints via the Event Horizon Telescope for Black Hole Solutions with Dark Matter under the Generalized Uncertainty Principle Minimal Length Scale Effect}},}\ }\href {\doibase 10.1002/andp.202300390} {\bibfield  {journal} {\bibinfo  {journal} {Annalen Phys.}\ }\textbf {\bibinfo {volume} {2023}},\ \bibinfo {pages} {2300390} (\bibinfo {year} {2023})},\ \Eprint {http://arxiv.org/abs/2309.07442} {arXiv:2309.07442 [gr-qc]} \BibitemShut {NoStop}%
\bibitem [{\citenamefont {Pantig}(2024)}]{Pantig:2024rmr}%
  \BibitemOpen
  \bibfield  {author} {\bibinfo {author} {\bibfnamefont {Reggie~C.}\ \bibnamefont {Pantig}},\ }\bibfield  {title} {\enquote {\bibinfo {title} {{Apparent and emergent dark matter around a Schwarzschild black hole}},}\ }\href {\doibase 10.1016/j.dark.2024.101550} {\bibfield  {journal} {\bibinfo  {journal} {Phys. Dark Univ.}\ }\textbf {\bibinfo {volume} {45}},\ \bibinfo {pages} {101550} (\bibinfo {year} {2024})},\ \Eprint {http://arxiv.org/abs/2405.07531} {arXiv:2405.07531 [gr-qc]} \BibitemShut {NoStop}%
\bibitem [{\citenamefont {Nozari}\ \emph {et~al.}(2025)\citenamefont {Nozari}, \citenamefont {Saghafi},\ and\ \citenamefont {Hassani}}]{Nozari:2024vxp}%
  \BibitemOpen
  \bibfield  {author} {\bibinfo {author} {\bibfnamefont {Kourosh}\ \bibnamefont {Nozari}}, \bibinfo {author} {\bibfnamefont {Sara}\ \bibnamefont {Saghafi}}, \ and\ \bibinfo {author} {\bibfnamefont {Mohammad}\ \bibnamefont {Hassani}},\ }\bibfield  {title} {\enquote {\bibinfo {title} {{Accretion onto a charged black hole in consistent 4D Einstein-Gauss-Bonnet gravity}},}\ }\href {\doibase 10.1016/j.jheap.2024.12.004} {\bibfield  {journal} {\bibinfo  {journal} {JHEAp}\ }\textbf {\bibinfo {volume} {45}},\ \bibinfo {pages} {214--230} (\bibinfo {year} {2025})},\ \Eprint {http://arxiv.org/abs/2412.07814} {arXiv:2412.07814 [gr-qc]} \BibitemShut {NoStop}%
\bibitem [{\citenamefont {Sui}\ \emph {et~al.}(2024)\citenamefont {Sui}, \citenamefont {Wang},\ and\ \citenamefont {Guo}}]{Sui:2023tje}%
  \BibitemOpen
  \bibfield  {author} {\bibinfo {author} {\bibfnamefont {Tao-Tao}\ \bibnamefont {Sui}}, \bibinfo {author} {\bibfnamefont {Zi-Liang}\ \bibnamefont {Wang}}, \ and\ \bibinfo {author} {\bibfnamefont {Wen-Di}\ \bibnamefont {Guo}},\ }\bibfield  {title} {\enquote {\bibinfo {title} {{The effect of scalar hair on the charged black hole with the images from accretions disk}},}\ }\href {\doibase 10.1140/epjc/s10052-024-12807-5} {\bibfield  {journal} {\bibinfo  {journal} {Eur. Phys. J. C}\ }\textbf {\bibinfo {volume} {84}},\ \bibinfo {pages} {441} (\bibinfo {year} {2024})},\ \Eprint {http://arxiv.org/abs/2311.10946} {arXiv:2311.10946 [gr-qc]} \BibitemShut {NoStop}%
\bibitem [{\citenamefont {Zare}\ \emph {et~al.}(2024)\citenamefont {Zare}, \citenamefont {Nieto}, \citenamefont {Feng}, \citenamefont {Dong},\ and\ \citenamefont {Hassanabadi}}]{Zare:2024dtf}%
  \BibitemOpen
  \bibfield  {author} {\bibinfo {author} {\bibfnamefont {Soroush}\ \bibnamefont {Zare}}, \bibinfo {author} {\bibfnamefont {Luis~M.}\ \bibnamefont {Nieto}}, \bibinfo {author} {\bibfnamefont {Xing-Hui}\ \bibnamefont {Feng}}, \bibinfo {author} {\bibfnamefont {Shi-Hai}\ \bibnamefont {Dong}}, \ and\ \bibinfo {author} {\bibfnamefont {Hassan}\ \bibnamefont {Hassanabadi}},\ }\bibfield  {title} {\enquote {\bibinfo {title} {{Shadows, rings and optical appearance of a magnetically charged regular black hole illuminated by various accretion disks}},}\ }\href {\doibase 10.1088/1475-7516/2024/08/041} {\bibfield  {journal} {\bibinfo  {journal} {JCAP}\ }\textbf {\bibinfo {volume} {08}},\ \bibinfo {pages} {041} (\bibinfo {year} {2024})},\ \Eprint {http://arxiv.org/abs/2406.07300} {arXiv:2406.07300 [astro-ph.HE]} \BibitemShut {NoStop}%
\bibitem [{\citenamefont {Pantig}\ and\ \citenamefont {\"Ovg\"un}(2023)}]{Pantig:2022sjb}%
  \BibitemOpen
  \bibfield  {author} {\bibinfo {author} {\bibfnamefont {Reggie~C.}\ \bibnamefont {Pantig}}\ and\ \bibinfo {author} {\bibfnamefont {Ali}\ \bibnamefont {\"Ovg\"un}},\ }\bibfield  {title} {\enquote {\bibinfo {title} {{Black Hole in Quantum Wave Dark Matter}},}\ }\href {\doibase 10.1002/prop.202200164} {\bibfield  {journal} {\bibinfo  {journal} {Fortsch. Phys.}\ }\textbf {\bibinfo {volume} {71}},\ \bibinfo {pages} {2200164} (\bibinfo {year} {2023})},\ \Eprint {http://arxiv.org/abs/2210.00523} {arXiv:2210.00523 [gr-qc]} \BibitemShut {NoStop}%
\bibitem [{\citenamefont {Sucu}\ and\ \citenamefont {\"Ovg\"un}(2024)}]{Sucu:2024xck}%
  \BibitemOpen
  \bibfield  {author} {\bibinfo {author} {\bibfnamefont {Erdem}\ \bibnamefont {Sucu}}\ and\ \bibinfo {author} {\bibfnamefont {Ali}\ \bibnamefont {\"Ovg\"un}},\ }\bibfield  {title} {\enquote {\bibinfo {title} {{The effect of quark\textendash{}antiquark confinement on the deflection angle by the NED black hole}},}\ }\href {\doibase 10.1016/j.dark.2024.101446} {\bibfield  {journal} {\bibinfo  {journal} {Phys. Dark Univ.}\ }\textbf {\bibinfo {volume} {44}},\ \bibinfo {pages} {101446} (\bibinfo {year} {2024})},\ \Eprint {http://arxiv.org/abs/2403.07044} {arXiv:2403.07044 [gr-qc]} \BibitemShut {NoStop}%
\bibitem [{\citenamefont {Ara\'ujo~Filho}(2025{\natexlab{a}})}]{AraujoFilho:2024xhm}%
  \BibitemOpen
  \bibfield  {author} {\bibinfo {author} {\bibfnamefont {A.~A.}\ \bibnamefont {Ara\'ujo~Filho}},\ }\bibfield  {title} {\enquote {\bibinfo {title} {{Analysis of a nonlinear electromagnetic generalization of the Reissner\textendash{}Nordstr\"om black hole}},}\ }\href {\doibase 10.1140/epjc/s10052-025-14182-1} {\bibfield  {journal} {\bibinfo  {journal} {Eur. Phys. J. C}\ }\textbf {\bibinfo {volume} {85}},\ \bibinfo {pages} {454} (\bibinfo {year} {2025}{\natexlab{a}})},\ \Eprint {http://arxiv.org/abs/2410.12060} {arXiv:2410.12060 [gr-qc]} \BibitemShut {NoStop}%
\bibitem [{\citenamefont {Ara\'ujo~Filho}(2025{\natexlab{b}})}]{AraujoFilho:2024lsi}%
  \BibitemOpen
  \bibfield  {author} {\bibinfo {author} {\bibfnamefont {A.~A.}\ \bibnamefont {Ara\'ujo~Filho}},\ }\bibfield  {title} {\enquote {\bibinfo {title} {{Remarks on a nonlinear electromagnetic extension in AdS Reissner-Nordstr\"om spacetime}},}\ }\href {\doibase 10.1088/1475-7516/2025/01/072} {\bibfield  {journal} {\bibinfo  {journal} {JCAP}\ }\textbf {\bibinfo {volume} {01}},\ \bibinfo {pages} {072} (\bibinfo {year} {2025}{\natexlab{b}})},\ \Eprint {http://arxiv.org/abs/2410.23165} {arXiv:2410.23165 [gr-qc]} \BibitemShut {NoStop}%
\bibitem [{\citenamefont {Uniyal}\ \emph {et~al.}(2024{\natexlab{a}})\citenamefont {Uniyal}, \citenamefont {Chakrabarti}, \citenamefont {Pantig},\ and\ \citenamefont {\"Ovg\"un}}]{Uniyal:2023inx}%
  \BibitemOpen
  \bibfield  {author} {\bibinfo {author} {\bibfnamefont {Akhil}\ \bibnamefont {Uniyal}}, \bibinfo {author} {\bibfnamefont {Sayan}\ \bibnamefont {Chakrabarti}}, \bibinfo {author} {\bibfnamefont {Reggie~C.}\ \bibnamefont {Pantig}}, \ and\ \bibinfo {author} {\bibfnamefont {Ali}\ \bibnamefont {\"Ovg\"un}},\ }\bibfield  {title} {\enquote {\bibinfo {title} {{Nonlinearly charged black holes: Shadow and thin-accretion disk}},}\ }\href {\doibase 10.1016/j.newast.2024.102249} {\bibfield  {journal} {\bibinfo  {journal} {New Astron.}\ }\textbf {\bibinfo {volume} {111}},\ \bibinfo {pages} {102249} (\bibinfo {year} {2024}{\natexlab{a}})},\ \Eprint {http://arxiv.org/abs/2303.07174} {arXiv:2303.07174 [gr-qc]} \BibitemShut {NoStop}%
\bibitem [{\citenamefont {Kumar~Walia}(2024)}]{KumarWalia:2024yxn}%
  \BibitemOpen
  \bibfield  {author} {\bibinfo {author} {\bibfnamefont {Rahul}\ \bibnamefont {Kumar~Walia}},\ }\bibfield  {title} {\enquote {\bibinfo {title} {{Exploring nonlinear electrodynamics theories: Shadows of regular black holes and horizonless ultracompact objects}},}\ }\href {\doibase 10.1103/PhysRevD.110.064058} {\bibfield  {journal} {\bibinfo  {journal} {Phys. Rev. D}\ }\textbf {\bibinfo {volume} {110}},\ \bibinfo {pages} {064058} (\bibinfo {year} {2024})},\ \Eprint {http://arxiv.org/abs/2409.13290} {arXiv:2409.13290 [gr-qc]} \BibitemShut {NoStop}%
\bibitem [{\citenamefont {Uniyal}\ \emph {et~al.}(2024{\natexlab{b}})\citenamefont {Uniyal}, \citenamefont {Chakrabarti}, \citenamefont {Fathi},\ and\ \citenamefont {\"Ovg\"un}}]{Uniyal:2023ahv}%
  \BibitemOpen
  \bibfield  {author} {\bibinfo {author} {\bibfnamefont {Akhil}\ \bibnamefont {Uniyal}}, \bibinfo {author} {\bibfnamefont {Sayan}\ \bibnamefont {Chakrabarti}}, \bibinfo {author} {\bibfnamefont {Mohsen}\ \bibnamefont {Fathi}}, \ and\ \bibinfo {author} {\bibfnamefont {Ali}\ \bibnamefont {\"Ovg\"un}},\ }\bibfield  {title} {\enquote {\bibinfo {title} {{Observational signatures: Shadow cast by the effective metric of photons for black holes with rational non-linear electrodynamics}},}\ }\href {\doibase 10.1016/j.aop.2024.169614} {\bibfield  {journal} {\bibinfo  {journal} {Annals Phys.}\ }\textbf {\bibinfo {volume} {462}},\ \bibinfo {pages} {169614} (\bibinfo {year} {2024}{\natexlab{b}})},\ \Eprint {http://arxiv.org/abs/2309.13680} {arXiv:2309.13680 [gr-qc]} \BibitemShut {NoStop}%
\bibitem [{\citenamefont {Lambiase}\ \emph {et~al.}(2024)\citenamefont {Lambiase}, \citenamefont {Pantig},\ and\ \citenamefont {\"Ovg\"un}}]{Lambiase:2024uzy}%
  \BibitemOpen
  \bibfield  {author} {\bibinfo {author} {\bibfnamefont {Gaetano}\ \bibnamefont {Lambiase}}, \bibinfo {author} {\bibfnamefont {Reggie~C.}\ \bibnamefont {Pantig}}, \ and\ \bibinfo {author} {\bibfnamefont {Ali}\ \bibnamefont {\"Ovg\"un}},\ }\bibfield  {title} {\enquote {\bibinfo {title} {{Weak field deflection angle and analytical parameter estimation of the Lorentz-violating Bumblebee parameter through the black hole shadow using EHT data}},}\ }\href {\doibase 10.1209/0295-5075/ad8d79} {\bibfield  {journal} {\bibinfo  {journal} {EPL}\ }\textbf {\bibinfo {volume} {148}},\ \bibinfo {pages} {49001} (\bibinfo {year} {2024})},\ \Eprint {http://arxiv.org/abs/2408.09620} {arXiv:2408.09620 [gr-qc]} \BibitemShut {NoStop}%
\bibitem [{\citenamefont {Pantig}\ \emph {et~al.}(2024)\citenamefont {Pantig}, \citenamefont {Kala}, \citenamefont {\"Ovg\"un},\ and\ \citenamefont {Lobos}}]{Pantig:2024ixc}%
  \BibitemOpen
  \bibfield  {author} {\bibinfo {author} {\bibfnamefont {Reggie~C.}\ \bibnamefont {Pantig}}, \bibinfo {author} {\bibfnamefont {Shubham}\ \bibnamefont {Kala}}, \bibinfo {author} {\bibfnamefont {Ali}\ \bibnamefont {\"Ovg\"un}}, \ and\ \bibinfo {author} {\bibfnamefont {Nikko John Leo~S.}\ \bibnamefont {Lobos}},\ }\href@noop {} {\enquote {\bibinfo {title} {{Testing black holes with cosmological constant in Einstein-bumblebee gravity through the black hole shadow using EHT data and deflection angle}},}\ } (\bibinfo {year} {2024}),\ \Eprint {http://arxiv.org/abs/2410.13661} {arXiv:2410.13661 [gr-qc]} \BibitemShut {NoStop}%
\bibitem [{\citenamefont {Pantig}(2025)}]{Pantig:2024kqy}%
  \BibitemOpen
  \bibfield  {author} {\bibinfo {author} {\bibfnamefont {Reggie~C.}\ \bibnamefont {Pantig}},\ }\bibfield  {title} {\enquote {\bibinfo {title} {{On the analytic generalization of particle deflection in the weak field regime and shadow size in light of EHT constraints for Schwarzschild-like black hole solutions}},}\ }\href {\doibase 10.1140/epjc/s10052-025-13766-1} {\bibfield  {journal} {\bibinfo  {journal} {Eur. Phys. J. C}\ }\textbf {\bibinfo {volume} {85}},\ \bibinfo {pages} {52} (\bibinfo {year} {2025})},\ \Eprint {http://arxiv.org/abs/2409.00476} {arXiv:2409.00476 [gr-qc]} \BibitemShut {NoStop}%
\bibitem [{\citenamefont {Li}\ \emph {et~al.}(2024)\citenamefont {Li}, \citenamefont {Zhang},\ and\ \citenamefont {Huang}}]{Li:2024owp}%
  \BibitemOpen
  \bibfield  {author} {\bibinfo {author} {\bibfnamefont {Hui-Ling}\ \bibnamefont {Li}}, \bibinfo {author} {\bibfnamefont {Miao}\ \bibnamefont {Zhang}}, \ and\ \bibinfo {author} {\bibfnamefont {Yu-Meng}\ \bibnamefont {Huang}},\ }\bibfield  {title} {\enquote {\bibinfo {title} {{The shadows of quintessence non-singular black hole}},}\ }\href {\doibase 10.1140/epjc/s10052-024-13194-7} {\bibfield  {journal} {\bibinfo  {journal} {Eur. Phys. J. C}\ }\textbf {\bibinfo {volume} {84}},\ \bibinfo {pages} {860} (\bibinfo {year} {2024})}\BibitemShut {NoStop}%
\bibitem [{\citenamefont {Khodadi}\ \emph {et~al.}(2024)\citenamefont {Khodadi}, \citenamefont {Vagnozzi},\ and\ \citenamefont {Firouzjaee}}]{Khodadi:2024ubi}%
  \BibitemOpen
  \bibfield  {author} {\bibinfo {author} {\bibfnamefont {Mohsen}\ \bibnamefont {Khodadi}}, \bibinfo {author} {\bibfnamefont {Sunny}\ \bibnamefont {Vagnozzi}}, \ and\ \bibinfo {author} {\bibfnamefont {Javad~T.}\ \bibnamefont {Firouzjaee}},\ }\bibfield  {title} {\enquote {\bibinfo {title} {{Event Horizon Telescope observations exclude compact objects in baseline mimetic gravity}},}\ }\href {\doibase 10.1038/s41598-024-78264-y} {\bibfield  {journal} {\bibinfo  {journal} {Sci. Rep.}\ }\textbf {\bibinfo {volume} {14}},\ \bibinfo {pages} {26932} (\bibinfo {year} {2024})},\ \Eprint {http://arxiv.org/abs/2408.03241} {arXiv:2408.03241 [gr-qc]} \BibitemShut {NoStop}%
\bibitem [{\citenamefont {Ali}\ \emph {et~al.}(2025)\citenamefont {Ali}, \citenamefont {Tiecheng}, \citenamefont {Babar},\ and\ \citenamefont {\"Ovg\"un}}]{Ali:2024cti}%
  \BibitemOpen
  \bibfield  {author} {\bibinfo {author} {\bibfnamefont {Riasat}\ \bibnamefont {Ali}}, \bibinfo {author} {\bibfnamefont {Xia}\ \bibnamefont {Tiecheng}}, \bibinfo {author} {\bibfnamefont {Rimsha}\ \bibnamefont {Babar}}, \ and\ \bibinfo {author} {\bibfnamefont {Ali}\ \bibnamefont {\"Ovg\"un}},\ }\bibfield  {title} {\enquote {\bibinfo {title} {{Exploring Light Deflection and Black Hole Shadows in Rastall Theory with Plasma Effects}},}\ }\href {\doibase 10.1007/s10773-025-05942-6} {\bibfield  {journal} {\bibinfo  {journal} {Int. J. Theor. Phys.}\ }\textbf {\bibinfo {volume} {64}},\ \bibinfo {pages} {75} (\bibinfo {year} {2025})},\ \Eprint {http://arxiv.org/abs/2402.07657} {arXiv:2402.07657 [gr-qc]} \BibitemShut {NoStop}%
\bibitem [{\citenamefont {Meng}\ \emph {et~al.}(2024)\citenamefont {Meng}, \citenamefont {Kuang}, \citenamefont {Wang}, \citenamefont {Wang},\ and\ \citenamefont {Wu}}]{Meng:2024puu}%
  \BibitemOpen
  \bibfield  {author} {\bibinfo {author} {\bibfnamefont {Yuan}\ \bibnamefont {Meng}}, \bibinfo {author} {\bibfnamefont {Xiao-Mei}\ \bibnamefont {Kuang}}, \bibinfo {author} {\bibfnamefont {Xi-Jing}\ \bibnamefont {Wang}}, \bibinfo {author} {\bibfnamefont {Bin}\ \bibnamefont {Wang}}, \ and\ \bibinfo {author} {\bibfnamefont {Jian-Pin}\ \bibnamefont {Wu}},\ }\bibfield  {title} {\enquote {\bibinfo {title} {{Images of hairy Reissner\textendash{}Nordstr\"om black hole illuminated by static accretions}},}\ }\href {\doibase 10.1140/epjc/s10052-024-12686-w} {\bibfield  {journal} {\bibinfo  {journal} {Eur. Phys. J. C}\ }\textbf {\bibinfo {volume} {84}},\ \bibinfo {pages} {305} (\bibinfo {year} {2024})},\ \Eprint {http://arxiv.org/abs/2401.05634} {arXiv:2401.05634 [gr-qc]} \BibitemShut {NoStop}%
\bibitem [{\citenamefont {Heidari}\ \emph {et~al.}(2024{\natexlab{a}})\citenamefont {Heidari}, \citenamefont {Hassanabadi}, \citenamefont {Filho},\ and\ \citenamefont {Kr̆\'\i{}z̆}}]{Heidari:2023egu}%
  \BibitemOpen
  \bibfield  {author} {\bibinfo {author} {\bibfnamefont {N.}~\bibnamefont {Heidari}}, \bibinfo {author} {\bibfnamefont {H.}~\bibnamefont {Hassanabadi}}, \bibinfo {author} {\bibfnamefont {A.~A.~Ara\'ujo}\ \bibnamefont {Filho}}, \ and\ \bibinfo {author} {\bibfnamefont {J.}~\bibnamefont {Kr̆\'\i{}z̆}},\ }\bibfield  {title} {\enquote {\bibinfo {title} {{Exploring non-commutativity as a perturbation in the Schwarzschild black hole: quasinormal modes, scattering, and shadows}},}\ }\href {\doibase 10.1140/epjc/s10052-024-12889-1} {\bibfield  {journal} {\bibinfo  {journal} {Eur. Phys. J. C}\ }\textbf {\bibinfo {volume} {84}},\ \bibinfo {pages} {566} (\bibinfo {year} {2024}{\natexlab{a}})},\ \Eprint {http://arxiv.org/abs/2308.03284} {arXiv:2308.03284 [gr-qc]} \BibitemShut {NoStop}%
\bibitem [{\citenamefont {Heidari}\ \emph {et~al.}(2024{\natexlab{b}})\citenamefont {Heidari}, \citenamefont {Hassanabadi}, \citenamefont {Filho}, \citenamefont {Kr̆\'\i{}z̆}, \citenamefont {Zare},\ and\ \citenamefont {Porf\'\i{}rio}}]{Heidari:2023bww}%
  \BibitemOpen
  \bibfield  {author} {\bibinfo {author} {\bibfnamefont {N.}~\bibnamefont {Heidari}}, \bibinfo {author} {\bibfnamefont {H.}~\bibnamefont {Hassanabadi}}, \bibinfo {author} {\bibfnamefont {A.~A.~Ara\'ujo}\ \bibnamefont {Filho}}, \bibinfo {author} {\bibfnamefont {J.}~\bibnamefont {Kr̆\'\i{}z̆}}, \bibinfo {author} {\bibfnamefont {S.}~\bibnamefont {Zare}}, \ and\ \bibinfo {author} {\bibfnamefont {P.~J.}\ \bibnamefont {Porf\'\i{}rio}},\ }\bibfield  {title} {\enquote {\bibinfo {title} {{Gravitational signatures of a non-commutative stable black hole}},}\ }\href {\doibase 10.1016/j.dark.2023.101382} {\bibfield  {journal} {\bibinfo  {journal} {Phys. Dark Univ.}\ }\textbf {\bibinfo {volume} {43}},\ \bibinfo {pages} {101382} (\bibinfo {year} {2024}{\natexlab{b}})},\ \Eprint {http://arxiv.org/abs/2305.06838} {arXiv:2305.06838 [gr-qc]} \BibitemShut {NoStop}%
\bibitem [{\citenamefont {Calz\`a}\ \emph {et~al.}(2025{\natexlab{a}})\citenamefont {Calz\`a}, \citenamefont {Pedrotti},\ and\ \citenamefont {Vagnozzi}}]{Calza:2024xdh}%
  \BibitemOpen
  \bibfield  {author} {\bibinfo {author} {\bibfnamefont {Marco}\ \bibnamefont {Calz\`a}}, \bibinfo {author} {\bibfnamefont {Davide}\ \bibnamefont {Pedrotti}}, \ and\ \bibinfo {author} {\bibfnamefont {Sunny}\ \bibnamefont {Vagnozzi}},\ }\bibfield  {title} {\enquote {\bibinfo {title} {{Primordial regular black holes as all the dark matter. II. Non-time-radial-symmetric and loop quantum gravity-inspired metrics}},}\ }\href {\doibase 10.1103/PhysRevD.111.024010} {\bibfield  {journal} {\bibinfo  {journal} {Phys. Rev. D}\ }\textbf {\bibinfo {volume} {111}},\ \bibinfo {pages} {024010} (\bibinfo {year} {2025}{\natexlab{a}})},\ \Eprint {http://arxiv.org/abs/2409.02807} {arXiv:2409.02807 [gr-qc]} \BibitemShut {NoStop}%
\bibitem [{\citenamefont {Calz\`a}\ \emph {et~al.}(2025{\natexlab{b}})\citenamefont {Calz\`a}, \citenamefont {Pedrotti},\ and\ \citenamefont {Vagnozzi}}]{Calza:2024fzo}%
  \BibitemOpen
  \bibfield  {author} {\bibinfo {author} {\bibfnamefont {Marco}\ \bibnamefont {Calz\`a}}, \bibinfo {author} {\bibfnamefont {Davide}\ \bibnamefont {Pedrotti}}, \ and\ \bibinfo {author} {\bibfnamefont {Sunny}\ \bibnamefont {Vagnozzi}},\ }\bibfield  {title} {\enquote {\bibinfo {title} {{Primordial regular black holes as all the dark matter. I. Time-radial-symmetric metrics}},}\ }\href {\doibase 10.1103/PhysRevD.111.024009} {\bibfield  {journal} {\bibinfo  {journal} {Phys. Rev. D}\ }\textbf {\bibinfo {volume} {111}},\ \bibinfo {pages} {024009} (\bibinfo {year} {2025}{\natexlab{b}})},\ \Eprint {http://arxiv.org/abs/2409.02804} {arXiv:2409.02804 [gr-qc]} \BibitemShut {NoStop}%
\bibitem [{\citenamefont {Bozza}(2002)}]{Bozza:2002zj}%
  \BibitemOpen
  \bibfield  {author} {\bibinfo {author} {\bibfnamefont {V.}~\bibnamefont {Bozza}},\ }\bibfield  {title} {\enquote {\bibinfo {title} {{Gravitational lensing in the strong field limit}},}\ }\href {\doibase 10.1103/PhysRevD.66.103001} {\bibfield  {journal} {\bibinfo  {journal} {Phys. Rev. D}\ }\textbf {\bibinfo {volume} {66}},\ \bibinfo {pages} {103001} (\bibinfo {year} {2002})},\ \Eprint {http://arxiv.org/abs/gr-qc/0208075} {arXiv:gr-qc/0208075} \BibitemShut {NoStop}%
\bibitem [{\citenamefont {Tsukamoto}(2017)}]{Tsukamoto:2016jzh}%
  \BibitemOpen
  \bibfield  {author} {\bibinfo {author} {\bibfnamefont {Naoki}\ \bibnamefont {Tsukamoto}},\ }\bibfield  {title} {\enquote {\bibinfo {title} {{Deflection angle in the strong deflection limit in a general asymptotically flat, static, spherically symmetric spacetime}},}\ }\href {\doibase 10.1103/PhysRevD.95.064035} {\bibfield  {journal} {\bibinfo  {journal} {Phys. Rev. D}\ }\textbf {\bibinfo {volume} {95}},\ \bibinfo {pages} {064035} (\bibinfo {year} {2017})},\ \Eprint {http://arxiv.org/abs/1612.08251} {arXiv:1612.08251 [gr-qc]} \BibitemShut {NoStop}%
\bibitem [{\citenamefont {Tsukamoto}\ and\ \citenamefont {Gong}(2017)}]{Tsukamoto:2016oca}%
  \BibitemOpen
  \bibfield  {author} {\bibinfo {author} {\bibfnamefont {Naoki}\ \bibnamefont {Tsukamoto}}\ and\ \bibinfo {author} {\bibfnamefont {Yungui}\ \bibnamefont {Gong}},\ }\bibfield  {title} {\enquote {\bibinfo {title} {{Retrolensing by a charged black hole}},}\ }\href {\doibase 10.1103/PhysRevD.95.064034} {\bibfield  {journal} {\bibinfo  {journal} {Phys. Rev. D}\ }\textbf {\bibinfo {volume} {95}},\ \bibinfo {pages} {064034} (\bibinfo {year} {2017})},\ \Eprint {http://arxiv.org/abs/1612.08250} {arXiv:1612.08250 [gr-qc]} \BibitemShut {NoStop}%
\bibitem [{\citenamefont {Fu}\ \emph {et~al.}(2021)\citenamefont {Fu}, \citenamefont {Zhao},\ and\ \citenamefont {Liu}}]{Fu:2021akc}%
  \BibitemOpen
  \bibfield  {author} {\bibinfo {author} {\bibfnamefont {Qi-Ming}\ \bibnamefont {Fu}}, \bibinfo {author} {\bibfnamefont {Li}~\bibnamefont {Zhao}}, \ and\ \bibinfo {author} {\bibfnamefont {Yu-Xiao}\ \bibnamefont {Liu}},\ }\bibfield  {title} {\enquote {\bibinfo {title} {{Weak deflection angle by electrically and magnetically charged black holes from nonlinear electrodynamics}},}\ }\href {\doibase 10.1103/PhysRevD.104.024033} {\bibfield  {journal} {\bibinfo  {journal} {Phys. Rev. D}\ }\textbf {\bibinfo {volume} {104}},\ \bibinfo {pages} {024033} (\bibinfo {year} {2021})},\ \Eprint {http://arxiv.org/abs/2101.08409} {arXiv:2101.08409 [gr-qc]} \BibitemShut {NoStop}%
\bibitem [{\citenamefont {Soares}\ \emph {et~al.}(2023{\natexlab{a}})\citenamefont {Soares}, \citenamefont {Vit\'oria},\ and\ \citenamefont {Pereira}}]{Soares:2023err}%
  \BibitemOpen
  \bibfield  {author} {\bibinfo {author} {\bibfnamefont {A.~R.}\ \bibnamefont {Soares}}, \bibinfo {author} {\bibfnamefont {R.~L.~L.}\ \bibnamefont {Vit\'oria}}, \ and\ \bibinfo {author} {\bibfnamefont {C.~F.~S.}\ \bibnamefont {Pereira}},\ }\bibfield  {title} {\enquote {\bibinfo {title} {{Gravitational lensing in a topologically charged Eddington-inspired Born\textendash{}Infeld spacetime}},}\ }\href {\doibase 10.1140/epjc/s10052-023-12071-z} {\bibfield  {journal} {\bibinfo  {journal} {Eur. Phys. J. C}\ }\textbf {\bibinfo {volume} {83}},\ \bibinfo {pages} {903} (\bibinfo {year} {2023}{\natexlab{a}})},\ \Eprint {http://arxiv.org/abs/2305.11105} {arXiv:2305.11105 [gr-qc]} \BibitemShut {NoStop}%
\bibitem [{\citenamefont {Soares}\ \emph {et~al.}(2023{\natexlab{b}})\citenamefont {Soares}, \citenamefont {Pereira}, \citenamefont {Vit\'oria},\ and\ \citenamefont {Rocha}}]{Soares:2023uup}%
  \BibitemOpen
  \bibfield  {author} {\bibinfo {author} {\bibfnamefont {A.~R.}\ \bibnamefont {Soares}}, \bibinfo {author} {\bibfnamefont {C.~F.~S.}\ \bibnamefont {Pereira}}, \bibinfo {author} {\bibfnamefont {R.~L.~L.}\ \bibnamefont {Vit\'oria}}, \ and\ \bibinfo {author} {\bibfnamefont {Erick~Melo}\ \bibnamefont {Rocha}},\ }\bibfield  {title} {\enquote {\bibinfo {title} {{Holonomy corrected Schwarzschild black hole lensing}},}\ }\href {\doibase 10.1103/PhysRevD.108.124024} {\bibfield  {journal} {\bibinfo  {journal} {Phys. Rev. D}\ }\textbf {\bibinfo {volume} {108}},\ \bibinfo {pages} {124024} (\bibinfo {year} {2023}{\natexlab{b}})},\ \Eprint {http://arxiv.org/abs/2309.05106} {arXiv:2309.05106 [gr-qc]} \BibitemShut {NoStop}%
\bibitem [{\citenamefont {He}\ \emph {et~al.}(2022)\citenamefont {He}, \citenamefont {Tao}, \citenamefont {Wang}, \citenamefont {Xue},\ and\ \citenamefont {Zhang}}]{He:2022opa}%
  \BibitemOpen
  \bibfield  {author} {\bibinfo {author} {\bibfnamefont {Aoyun}\ \bibnamefont {He}}, \bibinfo {author} {\bibfnamefont {Jun}\ \bibnamefont {Tao}}, \bibinfo {author} {\bibfnamefont {Peng}\ \bibnamefont {Wang}}, \bibinfo {author} {\bibfnamefont {Yadong}\ \bibnamefont {Xue}}, \ and\ \bibinfo {author} {\bibfnamefont {Lingkai}\ \bibnamefont {Zhang}},\ }\bibfield  {title} {\enquote {\bibinfo {title} {{Effects of Born\textendash{}Infeld electrodynamics on black hole shadows}},}\ }\href {\doibase 10.1140/epjc/s10052-022-10637-x} {\bibfield  {journal} {\bibinfo  {journal} {Eur. Phys. J. C}\ }\textbf {\bibinfo {volume} {82}},\ \bibinfo {pages} {683} (\bibinfo {year} {2022})},\ \Eprint {http://arxiv.org/abs/2205.12779} {arXiv:2205.12779 [gr-qc]} \BibitemShut {NoStop}%
\bibitem [{\citenamefont {Allahyari}\ \emph {et~al.}(2020)\citenamefont {Allahyari}, \citenamefont {Khodadi}, \citenamefont {Vagnozzi},\ and\ \citenamefont {Mota}}]{Allahyari:2019jqz}%
  \BibitemOpen
  \bibfield  {author} {\bibinfo {author} {\bibfnamefont {Alireza}\ \bibnamefont {Allahyari}}, \bibinfo {author} {\bibfnamefont {Mohsen}\ \bibnamefont {Khodadi}}, \bibinfo {author} {\bibfnamefont {Sunny}\ \bibnamefont {Vagnozzi}}, \ and\ \bibinfo {author} {\bibfnamefont {David~F.}\ \bibnamefont {Mota}},\ }\bibfield  {title} {\enquote {\bibinfo {title} {{Magnetically charged black holes from non-linear electrodynamics and the Event Horizon Telescope}},}\ }\href {\doibase 10.1088/1475-7516/2020/02/003} {\bibfield  {journal} {\bibinfo  {journal} {JCAP}\ }\textbf {\bibinfo {volume} {02}},\ \bibinfo {pages} {003} (\bibinfo {year} {2020})},\ \Eprint {http://arxiv.org/abs/1912.08231} {arXiv:1912.08231 [gr-qc]} \BibitemShut {NoStop}%
\bibitem [{\citenamefont {Gullu}\ and\ \citenamefont {Mazharimousavi}(2021)}]{Gullu:2020qni}%
  \BibitemOpen
  \bibfield  {author} {\bibinfo {author} {\bibfnamefont {Ibrahim}\ \bibnamefont {Gullu}}\ and\ \bibinfo {author} {\bibfnamefont {S.~Habib}\ \bibnamefont {Mazharimousavi}},\ }\bibfield  {title} {\enquote {\bibinfo {title} {{Black holes in double-Logarithmic nonlinear electrodynamics}},}\ }\href {\doibase 10.1088/1402-4896/ac098f} {\bibfield  {journal} {\bibinfo  {journal} {Phys. Scripta}\ }\textbf {\bibinfo {volume} {96}},\ \bibinfo {pages} {095213} (\bibinfo {year} {2021})},\ \Eprint {http://arxiv.org/abs/2010.04603} {arXiv:2010.04603 [gr-qc]} \BibitemShut {NoStop}%
\bibitem [{\citenamefont {Kruglov}(2018)}]{Kruglov:2018rrm}%
  \BibitemOpen
  \bibfield  {author} {\bibinfo {author} {\bibfnamefont {S.~I.}\ \bibnamefont {Kruglov}},\ }\bibfield  {title} {\enquote {\bibinfo {title} {{Magnetically charged black hole in framework of nonlinear electrodynamics model}},}\ }\href {\doibase 10.1142/S0217751X18500239} {\bibfield  {journal} {\bibinfo  {journal} {Int. J. Mod. Phys. A}\ }\textbf {\bibinfo {volume} {33}},\ \bibinfo {pages} {1850023} (\bibinfo {year} {2018})},\ \Eprint {http://arxiv.org/abs/1803.02191} {arXiv:1803.02191 [physics.gen-ph]} \BibitemShut {NoStop}%
\bibitem [{\citenamefont {Ma}(2015)}]{Ma:2015gpa}%
  \BibitemOpen
  \bibfield  {author} {\bibinfo {author} {\bibfnamefont {Meng-Sen}\ \bibnamefont {Ma}},\ }\bibfield  {title} {\enquote {\bibinfo {title} {{Magnetically charged regular black hole in a model of nonlinear electrodynamics}},}\ }\href {\doibase 10.1016/j.aop.2015.08.028} {\bibfield  {journal} {\bibinfo  {journal} {Annals Phys.}\ }\textbf {\bibinfo {volume} {362}},\ \bibinfo {pages} {529--537} (\bibinfo {year} {2015})},\ \Eprint {http://arxiv.org/abs/1509.05580} {arXiv:1509.05580 [gr-qc]} \BibitemShut {NoStop}%
\bibitem [{\citenamefont {Dymnikova}\ and\ \citenamefont {Galaktionov}(2021)}]{Dymnikova:2021dqq}%
  \BibitemOpen
  \bibfield  {author} {\bibinfo {author} {\bibfnamefont {Irina}\ \bibnamefont {Dymnikova}}\ and\ \bibinfo {author} {\bibfnamefont {Evgeny}\ \bibnamefont {Galaktionov}},\ }\bibfield  {title} {\enquote {\bibinfo {title} {{Regular electrically charged objects in Nonlinear Electrodynamics coupled to Gravity}},}\ }\href {\doibase 10.1088/1742-6596/2103/1/012078} {\bibfield  {journal} {\bibinfo  {journal} {J. Phys. Conf. Ser.}\ }\textbf {\bibinfo {volume} {2103}},\ \bibinfo {pages} {012078} (\bibinfo {year} {2021})}\BibitemShut {NoStop}%
\bibitem [{\citenamefont {Mazharimousavi}(2024)}]{Mazharimousavi:2023omd}%
  \BibitemOpen
  \bibfield  {author} {\bibinfo {author} {\bibfnamefont {S.~Habib}\ \bibnamefont {Mazharimousavi}},\ }\bibfield  {title} {\enquote {\bibinfo {title} {{Confinement and nonlinear electrodynamics: Asymptotic Schwarzschild charged black hole}},}\ }\href {\doibase 10.1016/j.dark.2023.101413} {\bibfield  {journal} {\bibinfo  {journal} {Phys. Dark Univ.}\ }\textbf {\bibinfo {volume} {43}},\ \bibinfo {pages} {101413} (\bibinfo {year} {2024})}\BibitemShut {NoStop}%
\bibitem [{\citenamefont {Carmo}(1976)}]{carmo}%
  \BibitemOpen
  \bibfield  {author} {\bibinfo {author} {\bibfnamefont {M.~P.~Do}\ \bibnamefont {Carmo}},\ }\href@noop {} {\emph {\bibinfo {title} {Differential Geometry of Curves and Surfaces}}}\ (\bibinfo  {publisher} {Prentice-Hall},\ \bibinfo {address} {New Jersey},\ \bibinfo {year} {1976})\BibitemShut {NoStop}%
\bibitem [{\citenamefont {Gibbons}\ and\ \citenamefont {Werner}(2008)}]{Gibbons:2008rj}%
  \BibitemOpen
  \bibfield  {author} {\bibinfo {author} {\bibfnamefont {G.~W.}\ \bibnamefont {Gibbons}}\ and\ \bibinfo {author} {\bibfnamefont {M.~C.}\ \bibnamefont {Werner}},\ }\bibfield  {title} {\enquote {\bibinfo {title} {{Applications of the Gauss-Bonnet theorem to gravitational lensing}},}\ }\href {\doibase 10.1088/0264-9381/25/23/235009} {\bibfield  {journal} {\bibinfo  {journal} {Class. Quant. Grav.}\ }\textbf {\bibinfo {volume} {25}},\ \bibinfo {pages} {235009} (\bibinfo {year} {2008})},\ \Eprint {http://arxiv.org/abs/0807.0854} {arXiv:0807.0854 [gr-qc]} \BibitemShut {NoStop}%
\bibitem [{\citenamefont {Li}\ \emph {et~al.}(2020)\citenamefont {Li}, \citenamefont {He},\ and\ \citenamefont {Zhou}}]{Li:2019vhp}%
  \BibitemOpen
  \bibfield  {author} {\bibinfo {author} {\bibfnamefont {Zonghai}\ \bibnamefont {Li}}, \bibinfo {author} {\bibfnamefont {Guansheng}\ \bibnamefont {He}}, \ and\ \bibinfo {author} {\bibfnamefont {Tao}\ \bibnamefont {Zhou}},\ }\bibfield  {title} {\enquote {\bibinfo {title} {{Gravitational deflection of relativistic massive particles by wormholes}},}\ }\href {\doibase 10.1103/PhysRevD.101.044001} {\bibfield  {journal} {\bibinfo  {journal} {Phys. Rev. D}\ }\textbf {\bibinfo {volume} {101}},\ \bibinfo {pages} {044001} (\bibinfo {year} {2020})},\ \Eprint {http://arxiv.org/abs/1908.01647} {arXiv:1908.01647 [gr-qc]} \BibitemShut {NoStop}%
\bibitem [{\citenamefont {Will}(2018)}]{Will:2018bme}%
  \BibitemOpen
  \bibfield  {author} {\bibinfo {author} {\bibfnamefont {Clifford~M.}\ \bibnamefont {Will}},\ }\href@noop {} {\emph {\bibinfo {title} {{Theory and Experiment in Gravitational Physics}}}}\ (\bibinfo  {publisher} {Cambridge University Press},\ \bibinfo {year} {2018})\BibitemShut {NoStop}%
\bibitem [{\citenamefont {Fil'chenkov}\ and\ \citenamefont {Laptev}(2018)}]{Filchenkov:2018wom}%
  \BibitemOpen
  \bibfield  {author} {\bibinfo {author} {\bibfnamefont {M.~L.}\ \bibnamefont {Fil'chenkov}}\ and\ \bibinfo {author} {\bibfnamefont {Yu.~P.}\ \bibnamefont {Laptev}},\ }\bibfield  {title} {\enquote {\bibinfo {title} {{Eddington\textquoteright{}s Prediction for General Relativity Revisited}},}\ }\href {\doibase 10.1134/S0202289318040084} {\bibfield  {journal} {\bibinfo  {journal} {Grav. Cosmol.}\ }\textbf {\bibinfo {volume} {24}},\ \bibinfo {pages} {371--374} (\bibinfo {year} {2018})}\BibitemShut {NoStop}%
\bibitem [{\citenamefont {Perlick}\ and\ \citenamefont {Tsupko}(2022)}]{Perlick:2021aok}%
  \BibitemOpen
  \bibfield  {author} {\bibinfo {author} {\bibfnamefont {Volker}\ \bibnamefont {Perlick}}\ and\ \bibinfo {author} {\bibfnamefont {Oleg~Yu.}\ \bibnamefont {Tsupko}},\ }\bibfield  {title} {\enquote {\bibinfo {title} {{Calculating black hole shadows: Review of analytical studies}},}\ }\href {\doibase 10.1016/j.physrep.2021.10.004} {\bibfield  {journal} {\bibinfo  {journal} {Phys. Rept.}\ }\textbf {\bibinfo {volume} {947}},\ \bibinfo {pages} {1--39} (\bibinfo {year} {2022})},\ \Eprint {http://arxiv.org/abs/2105.07101} {arXiv:2105.07101 [gr-qc]} \BibitemShut {NoStop}%
\bibitem [{\citenamefont {Perlick}\ \emph {et~al.}(2015)\citenamefont {Perlick}, \citenamefont {Tsupko},\ and\ \citenamefont {Bisnovatyi-Kogan}}]{Perlick:2015vta}%
  \BibitemOpen
  \bibfield  {author} {\bibinfo {author} {\bibfnamefont {Volker}\ \bibnamefont {Perlick}}, \bibinfo {author} {\bibfnamefont {Oleg~Yu.}\ \bibnamefont {Tsupko}}, \ and\ \bibinfo {author} {\bibfnamefont {Gennady~S.}\ \bibnamefont {Bisnovatyi-Kogan}},\ }\bibfield  {title} {\enquote {\bibinfo {title} {{Influence of a plasma on the shadow of a spherically symmetric black hole}},}\ }\href {\doibase 10.1103/PhysRevD.92.104031} {\bibfield  {journal} {\bibinfo  {journal} {Phys. Rev. D}\ }\textbf {\bibinfo {volume} {92}},\ \bibinfo {pages} {104031} (\bibinfo {year} {2015})},\ \Eprint {http://arxiv.org/abs/1507.04217} {arXiv:1507.04217 [gr-qc]} \BibitemShut {NoStop}%
\bibitem [{\citenamefont {Vagnozzi}\ \emph {et~al.}(2023)\citenamefont {Vagnozzi} \emph {et~al.}}]{Vagnozzi:2022moj}%
  \BibitemOpen
  \bibfield  {author} {\bibinfo {author} {\bibfnamefont {Sunny}\ \bibnamefont {Vagnozzi}} \emph {et~al.},\ }\bibfield  {title} {\enquote {\bibinfo {title} {{Horizon-scale tests of gravity theories and fundamental physics from the Event Horizon Telescope image of Sagittarius A}},}\ }\href {\doibase 10.1088/1361-6382/acd97b} {\bibfield  {journal} {\bibinfo  {journal} {Class. Quant. Grav.}\ }\textbf {\bibinfo {volume} {40}},\ \bibinfo {pages} {165007} (\bibinfo {year} {2023})},\ \Eprint {http://arxiv.org/abs/2205.07787} {arXiv:2205.07787 [gr-qc]} \BibitemShut {NoStop}%
\bibitem [{\citenamefont {Kocherlakota}\ \emph {et~al.}(2021)\citenamefont {Kocherlakota} \emph {et~al.}}]{EventHorizonTelescope:2021dqv}%
  \BibitemOpen
  \bibfield  {author} {\bibinfo {author} {\bibfnamefont {Prashant}\ \bibnamefont {Kocherlakota}} \emph {et~al.} (\bibinfo {collaboration} {Event Horizon Telescope}),\ }\bibfield  {title} {\enquote {\bibinfo {title} {{Constraints on black-hole charges with the 2017 EHT observations of M87*}},}\ }\href {\doibase 10.1103/PhysRevD.103.104047} {\bibfield  {journal} {\bibinfo  {journal} {Phys. Rev. D}\ }\textbf {\bibinfo {volume} {103}},\ \bibinfo {pages} {104047} (\bibinfo {year} {2021})},\ \Eprint {http://arxiv.org/abs/2105.09343} {arXiv:2105.09343 [gr-qc]} \BibitemShut {NoStop}%
\bibitem [{\citenamefont {Blackburn}\ \emph {et~al.}(2019)\citenamefont {Blackburn} \emph {et~al.}}]{Blackburn:2019bly}%
  \BibitemOpen
  \bibfield  {author} {\bibinfo {author} {\bibfnamefont {Lindy}\ \bibnamefont {Blackburn}} \emph {et~al.},\ }\href@noop {} {\enquote {\bibinfo {title} {{Studying Black Holes on Horizon Scales with VLBI Ground Arrays}},}\ } (\bibinfo {year} {2019}),\ \Eprint {http://arxiv.org/abs/1909.01411} {arXiv:1909.01411 [astro-ph.IM]} \BibitemShut {NoStop}%
\bibitem [{\citenamefont {Doeleman}\ \emph {et~al.}(2023)\citenamefont {Doeleman} \emph {et~al.}}]{Doeleman:2023kzg}%
  \BibitemOpen
  \bibfield  {author} {\bibinfo {author} {\bibfnamefont {Sheperd~S.}\ \bibnamefont {Doeleman}} \emph {et~al.},\ }\bibfield  {title} {\enquote {\bibinfo {title} {{Reference Array and Design Consideration for the Next-Generation Event Horizon Telescope}},}\ }\href {\doibase 10.3390/galaxies11050107} {\bibfield  {journal} {\bibinfo  {journal} {Galaxies}\ }\textbf {\bibinfo {volume} {11}},\ \bibinfo {pages} {107} (\bibinfo {year} {2023})},\ \Eprint {http://arxiv.org/abs/2306.08787} {arXiv:2306.08787 [astro-ph.IM]} \BibitemShut {NoStop}%
\bibitem [{\citenamefont {Raymond}\ \emph {et~al.}(2021)\citenamefont {Raymond} \emph {et~al.}}]{Raymond:2021syf}%
  \BibitemOpen
  \bibfield  {author} {\bibinfo {author} {\bibfnamefont {Alexander~W.}\ \bibnamefont {Raymond}} \emph {et~al.},\ }\bibfield  {title} {\enquote {\bibinfo {title} {{Evaluation of New Submillimeter VLBI Sites for the Event Horizon Telescope}},}\ }\href {\doibase 10.3847/1538-3881/abc3c3} {\bibfield  {journal} {\bibinfo  {journal} {Astron. J.}\ }\textbf {\bibinfo {volume} {253}},\ \bibinfo {pages} {5} (\bibinfo {year} {2021})}\BibitemShut {NoStop}%
\bibitem [{\citenamefont {Johnson}\ \emph {et~al.}(2020)\citenamefont {Johnson} \emph {et~al.}}]{Johnson:2019ljv}%
  \BibitemOpen
  \bibfield  {author} {\bibinfo {author} {\bibfnamefont {Michael~D.}\ \bibnamefont {Johnson}} \emph {et~al.},\ }\bibfield  {title} {\enquote {\bibinfo {title} {{Universal interferometric signatures of a black hole\textquoteright{}s photon ring}},}\ }\href {\doibase 10.1126/sciadv.aaz1310} {\bibfield  {journal} {\bibinfo  {journal} {Sci. Adv.}\ }\textbf {\bibinfo {volume} {6}},\ \bibinfo {pages} {eaaz1310} (\bibinfo {year} {2020})},\ \Eprint {http://arxiv.org/abs/1907.04329} {arXiv:1907.04329 [astro-ph.IM]} \BibitemShut {NoStop}%
\bibitem [{\citenamefont {Kardashev}\ \emph {et~al.}(2014)\citenamefont {Kardashev} \emph {et~al.}}]{Kardashev:2014sjq}%
  \BibitemOpen
  \bibfield  {author} {\bibinfo {author} {\bibfnamefont {N.~S.}\ \bibnamefont {Kardashev}} \emph {et~al.},\ }\bibfield  {title} {\enquote {\bibinfo {title} {{Review of scientific topics for the Millimetron space observatory}},}\ }\href {\doibase 10.3367/UFNr.0184.201412c.1319} {\bibfield  {journal} {\bibinfo  {journal} {Phys. Usp.}\ }\textbf {\bibinfo {volume} {57}},\ \bibinfo {pages} {1199} (\bibinfo {year} {2014})},\ \Eprint {http://arxiv.org/abs/1502.06071} {arXiv:1502.06071 [astro-ph.IM]} \BibitemShut {NoStop}%
\bibitem [{\citenamefont {Chael}\ \emph {et~al.}(2023)\citenamefont {Chael}, \citenamefont {Issaoun}, \citenamefont {Pesce}, \citenamefont {Johnson}, \citenamefont {Ricarte}, \citenamefont {Fromm},\ and\ \citenamefont {Mizuno}}]{Chael:2022meh}%
  \BibitemOpen
  \bibfield  {author} {\bibinfo {author} {\bibfnamefont {Andrew}\ \bibnamefont {Chael}}, \bibinfo {author} {\bibfnamefont {Sara}\ \bibnamefont {Issaoun}}, \bibinfo {author} {\bibfnamefont {Dominic~W.}\ \bibnamefont {Pesce}}, \bibinfo {author} {\bibfnamefont {Michael~D.}\ \bibnamefont {Johnson}}, \bibinfo {author} {\bibfnamefont {Angelo}\ \bibnamefont {Ricarte}}, \bibinfo {author} {\bibfnamefont {Christian~M.}\ \bibnamefont {Fromm}}, \ and\ \bibinfo {author} {\bibfnamefont {Yosuke}\ \bibnamefont {Mizuno}},\ }\bibfield  {title} {\enquote {\bibinfo {title} {{Multifrequency Black Hole Imaging for the Next-generation Event Horizon Telescope}},}\ }\href {\doibase 10.3847/1538-4357/acb7e4} {\bibfield  {journal} {\bibinfo  {journal} {Astrophys. J.}\ }\textbf {\bibinfo {volume} {945}},\ \bibinfo {pages} {40} (\bibinfo {year} {2023})},\ \Eprint {http://arxiv.org/abs/2210.12226} {arXiv:2210.12226 [astro-ph.HE]} \BibitemShut {NoStop}%
\end{thebibliography}%
\end{document}